\documentclass[conference]{IEEEtran}
\IEEEoverridecommandlockouts

\usepackage{cite}
\usepackage{amsmath,amssymb}
\usepackage{graphicx}
\usepackage{booktabs}
\usepackage{multirow}
\usepackage{url}
\usepackage{xcolor}
\usepackage[hidelinks]{hyperref}

\graphicspath{{figures/}}

\def\BibTeX{{\rm B\kern-.05em{\sc i\kern-.025em b}\kern-.08em T\kern-.1667em\lower.7ex\hbox{E}\kern-.125emX}}

\begin{document}

\title{Compared to What? A Human-Anchored Security Benchmark\\
for LLM-Generated Infrastructure-as-Code}

\author{\IEEEauthorblockN{Animesh Shaw}
\IEEEauthorblockA{animesh15b@iimk.edu.in}
}

\maketitle

\begin{abstract}
Large language models are increasingly used to author Infrastructure-as-Code
(IaC), where a single insecure default is provisioned directly into a production
cloud environment. Prior evaluations report absolute vulnerability counts for
model-generated IaC, but without a human reference they cannot say whether
models are actually \emph{worse} than the engineers they assist. We present
GenIaC-SecBench, a benchmark of 100 deployment scenarios stratified by
architectural complexity, evaluated across 12 model configurations spanning six
vendors and both open and closed weights, yielding 1{,}196 generated artifacts
scanned by three independent policy engines (Checkov, Trivy, KICS) at complete
coverage. Crucially, we scan a corpus of 634 human-authored IaC templates with
the identical toolchain, providing the first size-matched human security
baseline for this task.
We find that vulnerability density is strongly inverse to artifact size
(Spearman $\rho=-0.55$, $p<10^{-77}$), so unmatched comparisons measure artifact
size rather than security. Matched on declared-resource count, \emph{every}
model configuration exceeds the human baseline, spanning
$3.21\times$--$3.87\times$ human vulnerability density, and the gap widens as
tasks get simpler ($4.9\times$ at one resource, $1.4\times$ at twenty or more).
A majority of our scenarios prescribe a security state rather than specifying
function alone, so we report results stratified by prompt class: pooled, the gap
is $3.50\times$, and excluding every scenario that explicitly requests an
insecure configuration still leaves all configurations above the baseline
($2.4\times$--$4.2\times$). The corpus cannot, however, isolate
\emph{unprompted} default posture, and we say so.
We further decompose ``reasoning'' into three arms---standard generation,
prompt-engineered chain-of-thought, and the vendor's own extended-thinking
API---and show that vendor extended thinking significantly outperforms prompted
chain-of-thought ($-12.0\%$, $p=0.0013$), while prompted chain-of-thought alone
is indistinguishable from standard generation ($-1.3\%$, n.s.). Instrumenting
token usage reveals why the effect is bounded: extended thinking consumes under
$1\%$ of the output budget on this task class.
We also report two negative results: the intuitive hypothesis that more
deployable models are more vulnerable is unsupported ($r=0.158$, $p=0.625$), and
classical complete-case Friedman testing is \emph{uncomputable} on realistic
benchmark designs, motivating the Skillings--Mack statistic. All code, data, and
regeneration scripts are released.
\end{abstract}

\begin{IEEEkeywords}
infrastructure as code, large language models, software security, empirical
software engineering, static analysis, cloud security
\end{IEEEkeywords}

\section{Introduction}

Infrastructure-as-Code (IaC) tools such as Terraform, AWS CloudFormation, Azure
Resource Manager, and Kubernetes manifests declare cloud infrastructure as
version-controlled artifacts. A misconfiguration in an IaC template is not a
latent code defect: it is provisioned, and the resulting storage bucket,
security group, or IAM role is exposed exactly as written. Security smells in
IaC are both common and consequential~\cite{rahman2019seven,verdet2023exploring}.

Large language models now generate substantial volumes of such code. A body of
work has established that LLM-generated \emph{application} code contains
vulnerabilities at non-trivial rates~\cite{pearce2022asleep,sandoval2023lost,tihanyi2025secure},
and that developers using AI assistants may write less secure code while
believing the opposite~\cite{perry2023users}. Evaluations specific to IaC are
newer and thinner~\cite{vargas2026securityfirst}.

A recurring limitation unites these studies: they report absolute or normalized
vulnerability counts \emph{for models only}. Reporting that a model averages
eight findings per declared resource invites the immediate question
\emph{compared to what?} Without a human reference measured through the same
toolchain, such numbers cannot distinguish three very different worlds: models
being genuinely careless, models simply emitting more infrastructure per file,
or a scanner ruleset that flags any template heavily. This paper supplies the
missing anchor.

\medskip\noindent\textbf{Contributions.}
\begin{enumerate}
  \item \textbf{A size-matched human security baseline.} We scan 634
    human-authored IaC templates with the same three engines and show that
    density must be compared within resource-count strata. Matched, every model
    configuration sits at $3.21\times$--$3.87\times$ the human baseline
    (Section~\ref{sec:human}).
  \item \textbf{A three-arm decomposition of ``reasoning.''} Separating
    prompt-engineered chain-of-thought from vendor extended-thinking APIs shows
    the two are not interchangeable for security (Section~\ref{sec:reasoning}).
  \item \textbf{Direct measurement of reasoning engagement.} Extended thinking
    consumes ${<}1\%$ of output tokens on IaC, bounding the achievable effect
    (Section~\ref{sec:reasoning}).
  \item \textbf{A methodological correction for incomplete benchmark designs.}
    Complete-case Friedman retains zero usable blocks here; we adopt
    Skillings--Mack~\cite{skillings1981distribution} (Section~\ref{sec:omnibus}).
  \item \textbf{Calibration of an LLM judge against three human experts},
    delimiting where automated scenario review is trustworthy
    (Section~\ref{sec:human-eval}).
  \item \textbf{Two negative results}, reported as findings rather than omitted
    (Section~\ref{sec:negative}).
\end{enumerate}

\section{Background and Related Work}

\subsection{Security of LLM-generated code}
Pearce et al.~\cite{pearce2022asleep} found roughly 40\% of Copilot completions
in security-relevant contexts contained weaknesses. User studies show developers
with AI assistance produce less secure code while reporting higher
confidence~\cite{perry2023users,sandoval2023lost}. Large-scale comparisons across
models confirm that security varies substantially by
model~\cite{tihanyi2025secure}. Benchmarks such as
SecurityEval~\cite{siddiq2022securityeval} and correctness suites such as
EvalPlus~\cite{liu2023your} target application code, where the unit of analysis
is a function. IaC differs: the unit is a declarative resource graph, correctness
is schema validity, and the security question is about defaults rather than
control flow.

\subsection{IaC security}
Rahman et al.~\cite{rahman2019seven} catalogued seven recurring security smells
in IaC scripts; subsequent work extended detection across
languages~\cite{saavedra2023glitch} and studied practitioner
behaviour~\cite{verdet2023exploring,rahman2019systematic}. Policy engines
including Checkov~\cite{checkov}, Trivy~\cite{trivy}, and KICS~\cite{kics}
operationalize such rules, commonly mapped to CIS
Benchmarks~\cite{cis_benchmarks}. The closest prior evaluation of
\emph{generated} IaC is Vargas et al.~\cite{vargas2026securityfirst}, which
evaluates text-to-Terraform generation. Our study differs in scope---four IaC
formats rather than one, complexity stratification, three independent engines,
multi-rater human validation---and, most importantly, in providing a human
security baseline rather than model-only counts.

\subsection{Reasoning modes}
Chain-of-thought prompting improves performance on reasoning-heavy
tasks~\cite{wei2022chain,kojima2022large}, though recent evidence suggests gains
concentrate in mathematical and symbolic domains~\cite{sprague2025cot}. Vendors
now expose reasoning as a first-class API parameter with an explicit token
budget~\cite{openai2024o1,deepseek2025r1,anthropic2025thinking}. These are
distinct mechanisms: one changes the prompt, the other allocates computation.
To our knowledge no prior security evaluation separates them---and, as we report
in Section~\ref{sec:threats}, conflating them is easy to do accidentally.

\subsection{LLM-as-a-judge}
Model-based evaluation is widely used~\cite{zheng2023judging,gu2025survey} but
exhibits known biases, including preference for self-generated
content~\cite{panickssery2024llm}. We therefore treat our judge as a secondary
signal and calibrate it against three human experts.

\section{Methodology}

\subsection{Scenarios}
The benchmark comprises 100 natural-language deployment scenarios: 60
\emph{simple} (single-service tasks) and 40 \emph{complex} (multi-component
architectures spanning networking, identity, data, and observability). Scenarios
describe a deployment task; a majority additionally prescribe a security state,
either protective (``a \emph{private} bucket with versioning enabled'') or
permissive (``an EC2 instance with a public IP and \emph{open SSH port}'').
Section~\ref{sec:promptclass} classifies every scenario on this axis and reports
results stratified accordingly, because a finding on a prompt that \emph{asked
for} the insecure state measures instruction-following rather than an unprompted
default. Coverage spans AWS, Azure, GCP, and provider-agnostic Kubernetes across
Terraform HCL, CloudFormation, ARM, and Kubernetes manifests.

\subsection{Model configurations}
Table~\ref{tab:models} lists the 12 configurations. Beyond nine distinct models
across six vendors (Anthropic, Google, OpenAI, Meta, Mistral, Microsoft), three configurations isolate reasoning mode on a fixed base
model: \emph{standard}, \emph{-cot} (a chain-of-thought system-prompt suffix),
and \emph{-thinking} (the vendor extended-thinking API). All calls are stateless
with a fixed system prompt requesting code-only output.

\begin{table}[t]
\caption{Model configurations. The three Claude Opus 4.6 rows differ only in
reasoning mode, isolating that variable.}
\label{tab:models}
\centering
\small
\begin{tabular}{@{}lll@{}}
\toprule
Configuration & Vendor & Reasoning mode \\
\midrule
claude-opus-4-6           & Anthropic & none \\
claude-opus-4-6-cot       & Anthropic & prompted CoT \\
claude-opus-4-6-thinking  & Anthropic & extended thinking (adaptive) \\
claude-sonnet-4-6         & Anthropic & none \\
gpt-5                     & OpenAI    & none \\
gpt-5-thinking            & OpenAI    & \texttt{reasoning\_effort=high} \\
gpt-4o                    & OpenAI    & none \\
gemini-3.1-pro            & Google    & none \\
gemini-3.7-flash          & Google    & none \\
llama3 / mistral / phi3   & open       & none (local) \\
\bottomrule
\end{tabular}
\end{table}

\subsection{Generation protocol}
\label{sec:protocol}
We record the protocol verbatim, since generation details materially affect
security posture and are frequently underspecified in this
literature~\cite{ralph2021empirical}.

Every request is a single stateless API call: no conversation history, no
few-shot examples, no retrieval, and no shared memory between scenarios. The
system prompt establishes the role (``senior cloud infrastructure engineer''),
requests production-ready IaC, and constrains output to a single fenced code
block with no surrounding commentary. The user prompt is generated from the
scenario record as a target-format instruction (e.g.\ ``Write Terraform HCL code
for AWS to: \emph{[scenario text]}''). The system prompt does not mention
security, hardening, or compliance. Individual scenario texts, however, often
do: Section~\ref{sec:promptclass} shows that a majority prescribe a security
state in one direction or the other, which bounds what the resulting findings
can be read as measuring.

Sampling temperature is fixed at $0.2$ for standard and prompted-CoT arms.
Extended-thinking arms use the vendor default of $1.0$, which the API mandates
when reasoning is enabled---a confound we cannot eliminate without disabling the
feature under test, and which we note as a limitation. The chain-of-thought arm
appends a single instruction to the system prompt directing step-by-step
reasoning before the code block, and changes nothing else.

Output token ceilings are set well above the observed maximum: complex scenarios
have a measured median of 19{,}562 output tokens, and truncated responses
(\texttt{finish\_reason} indicating the limit was reached) are \emph{discarded}
rather than written, since a truncated template remains syntactically parseable
and would silently deflate both validity and finding counts. Four complex
scenarios could not be completed even at the model maximum of 128k output tokens
and are reported as missing.

Requests are retried on transient failures with exponential backoff. Per-request
token usage, including reasoning tokens, is logged for the analysis in
Section~\ref{sec:reasoning}.

\subsection{Validation and scanning}
Generated artifacts are schema-validated (\texttt{terraform validate},
\texttt{cfn-lint}, ARM parsing, \texttt{kubeconform}), then scanned by three
independent engines from three vendors---Checkov~\cite{checkov},
Trivy~\cite{trivy}, and KICS~\cite{kics}---giving a convergent-validity argument
no single ruleset provides. We report \emph{complete} coverage: 1{,}196 of
1{,}196 artifacts scanned by all three engines, verified by a machine-readable
coverage manifest regenerated on every run.

\subsection{The human reference corpus}
\label{sec:corpus}
We draw 634 human-authored IaC templates from three public repositories of
production-style and reference infrastructure. Files are filtered by structural
heuristics (root keys such as \texttt{Resources:} or \texttt{apiVersion:}) to
exclude non-IaC content. This corpus is scanned with the \emph{identical}
toolchain and configuration as the generated artifacts, so any ruleset bias
applies equally to both sides of the comparison.

\subsection{Metrics}
Our primary metric is \emph{vulnerability density}: total findings across all
three engines divided by the number of declared resources, obtained by parsing
each artifact's syntax tree. Density normalizes for the fact that a template
declaring fifty resources has more opportunity to be flagged than one declaring
two. Resource counts are taken from the AST parse rather than from scanner
output, which degrades silently when a file cannot be parsed
(Section~\ref{sec:threats}).

\subsection{Statistical procedure}
Models are compared within complexity strata using the Skillings--Mack
statistic~\cite{skillings1981distribution}, the generalization of the Friedman
test~\cite{friedman1937use} to incomplete block designs; Section~\ref{sec:omnibus}
explains why the classical test is unusable here. Post-hoc pairwise comparisons
use Wilcoxon signed-rank tests~\cite{wilcoxon1945individual} with Holm
correction~\cite{holm1979simple}, following Dem\v{s}ar~\cite{demsar2006statistical}.
Counts are additionally modelled with a negative binomial
GEE~\cite{liang1986longitudinal,hilbe2011negative} using
$\log(\textit{resource\_count})$ as an exposure offset, so coefficients are
per-resource incidence rate ratios. Negative binomial rather than Poisson is
required empirically: the observed variance-to-mean ratio is 130.0.
Human--model comparisons use Mann--Whitney $U$~\cite{mann1947test}; structural
distributions are compared with two-sample Kolmogorov--Smirnov
tests~\cite{massey1951kolmogorov}.

\section{Results}

\subsection{Corpus}
Generation yielded 1{,}196 of a possible 1{,}200 artifacts (99.7\%). Four
complex scenarios could not be generated within the model's maximum output
window even at 128k tokens---itself a finding about the scale of
architecture these prompts elicit. Scanning produced 38{,}803 findings
(Checkov 14{,}017; KICS 14{,}033; Trivy 10{,}753).

\subsection{Prompt prescriptiveness, and what findings can be read as}
\label{sec:promptclass}
A scanner finding on a prompt that explicitly requested the insecure state does
not measure the same thing as a finding on a prompt that did not. The first
measures instruction-following; only the second measures an unprompted default.
We therefore classify every scenario on this axis and report results stratified
by it.

Classification is deterministic and rule-based rather than model-assigned, since
a model-assigned label would itself require auditing. The rules are conservative
in the direction that weakens our own headline: ambiguous prompts are not counted
as insecure-prescriptive, so that class is a lower bound. The classifier, its
rule table, and the resulting labels are released with the dataset.

Of 100 scenarios, 52 prescribe an insecure state, 33 prescribe a protective one,
and 15 are functional-only. The imbalance is concentrated in the simple stratum
(43 of 60 insecure-prescriptive), which is by construction: those scenarios were
written as security decision points. \textbf{Findings inherit that imbalance.}
Insecure-prescriptive prompts account for $73.2\%$ of simple-stratum findings,
against $18.8\%$ in the complex stratum.

\textbf{Independent classification.} An external audit of the released artifacts
by Chauhan~\cite{iacguardv} classified the same scenarios under different rules,
obtaining 49 insecure-prescriptive, 41 secure-prescriptive, and 10
functional-only, and attributing $84.1\%$ of simple-stratum findings to the
insecure-prescriptive class. The two classifications agree on 75 of 100
scenarios. Disagreements concentrate on complex prompts that prescribe security
in \emph{both} directions---private endpoints alongside a public-facing load
balancer---where the two schemes tie-break oppositely. We report both rather than
adjudicate, since the spread is itself the honest measure of how sharp this
distinction is.

\begin{figure}[t]
\centering
\includegraphics[width=\columnwidth]{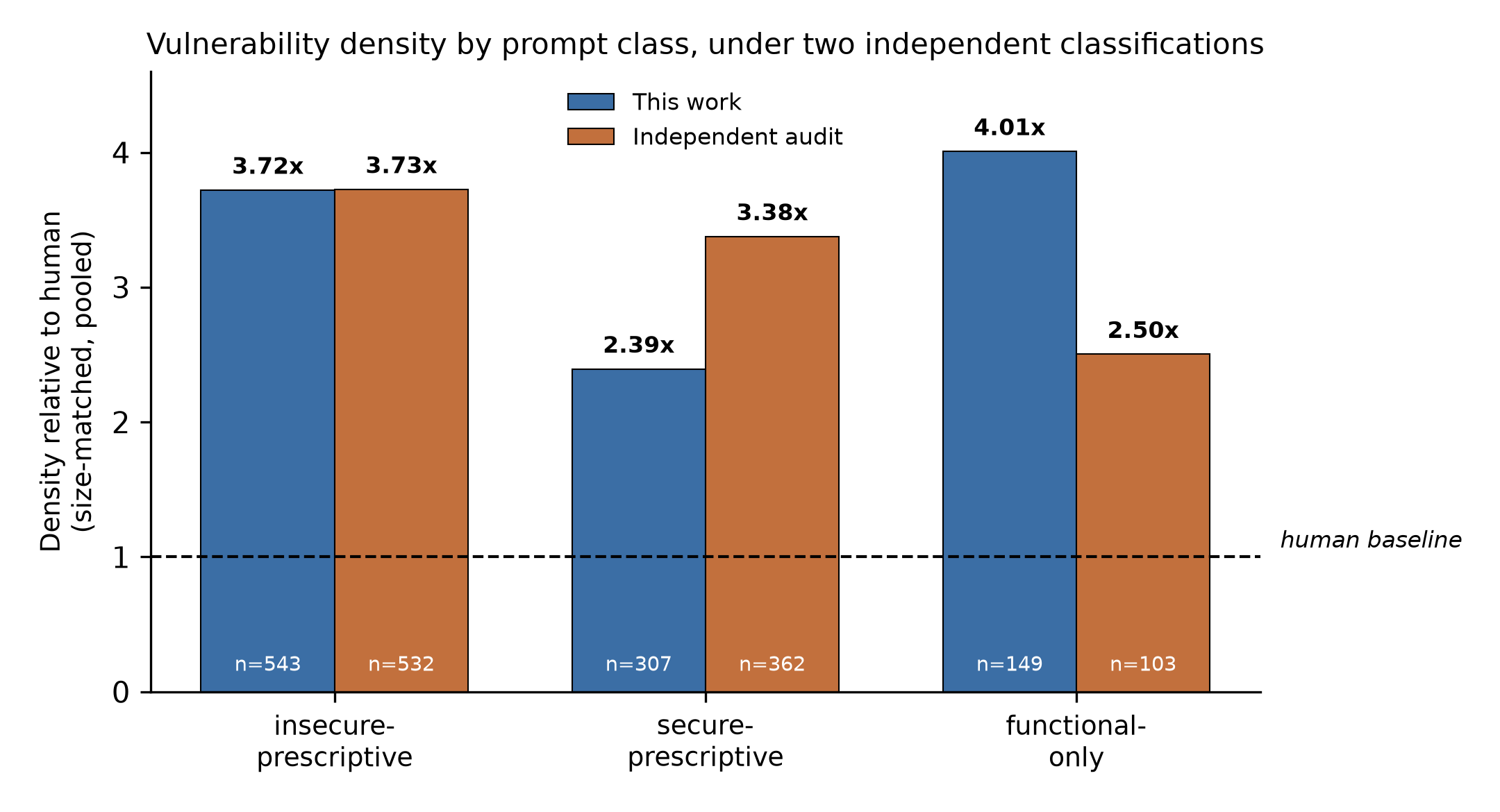}
\caption{Size-matched density by prompt class under two independent
classifications. Agreement on the insecure-prescriptive class and disagreement
on functional-only is why we decline to claim a model-default measurement.}
\label{fig:promptclass}
\end{figure}

Table~\ref{tab:promptclass} and Fig.~\ref{fig:promptclass} give size-matched
density by prompt class, pooled across configurations, under both schemes. The
figure makes the instability visible: the two schemes agree almost exactly on
the insecure-prescriptive class ($3.72\times$ versus $3.73\times$) and diverge
sharply on the functional-only class ($4.01\times$ versus $2.50\times$), which
is the stratum any claim about unprompted defaults would have to rest on.

\begin{table}[t]
\caption{Size-matched density relative to the human baseline, by prompt class,
pooled across configurations. The functional-only subset---the only stratum that
isolates unprompted defaults---is small and the two classifications disagree on
its direction.}
\label{tab:promptclass}
\centering
\small
\begin{tabular}{@{}lcccc@{}}
\toprule
 & \multicolumn{2}{c}{This work} & \multicolumn{2}{c}{Independent audit} \\
\cmidrule(lr){2-3}\cmidrule(lr){4-5}
Prompt class & ratio & $n$ & ratio & $n$ \\
\midrule
insecure-prescriptive & $3.72\times$ & 543 & $3.73\times$ & 532 \\
secure-prescriptive   & $2.39\times$ & 307 & $3.38\times$ & 362 \\
functional-only       & $4.01\times$ & 149 & $2.50\times$ & 103 \\
\midrule
All prompts (pooled)  & $3.50\times$ & 999 & $3.50\times$ & 999 \\
\bottomrule
\end{tabular}
\end{table}

\textbf{What this does and does not change.} The overall gap is robust: pooled,
it is $3.50\times$ under both schemes, and excluding insecure-prescriptive
prompts entirely leaves every configuration above the human baseline
($2.40\times$--$3.87\times$ under our classification, $2.48\times$--$4.21\times$
under the audit's). The per-bin result in Section~\ref{sec:matching} also
survives, and the single-resource bin strengthens from $4.90\times$ to
$5.45\times$ once insecure-prescriptive prompts are removed.

What weakens is the tightness of the cross-vendor band, which widens once the
insecure-prescriptive class is excluded, and any claim that this corpus measures
\emph{model-default} security posture. The functional-only subset is the only
stratum that could support such a claim, and at 103--149 artifacts spread over 12
configurations it is too small and too classification-sensitive to do so: the two
schemes place it at $4.01\times$ and $2.50\times$ respectively. Isolating default
posture from instruction-following requires a purpose-built functional-only
prompt set, which we identify as the most valuable extension of this work
(Section~\ref{sec:conclusion}). Reasoning-mode contrasts
(Section~\ref{sec:reasoning}) are unaffected by construction: they are paired
within-model over identical prompts, so prompt class is held constant and
cancels.

\subsection{Descriptive statistics}
\label{sec:descriptive}
\begin{figure*}[t]
\centering
\includegraphics[width=0.86\textwidth]{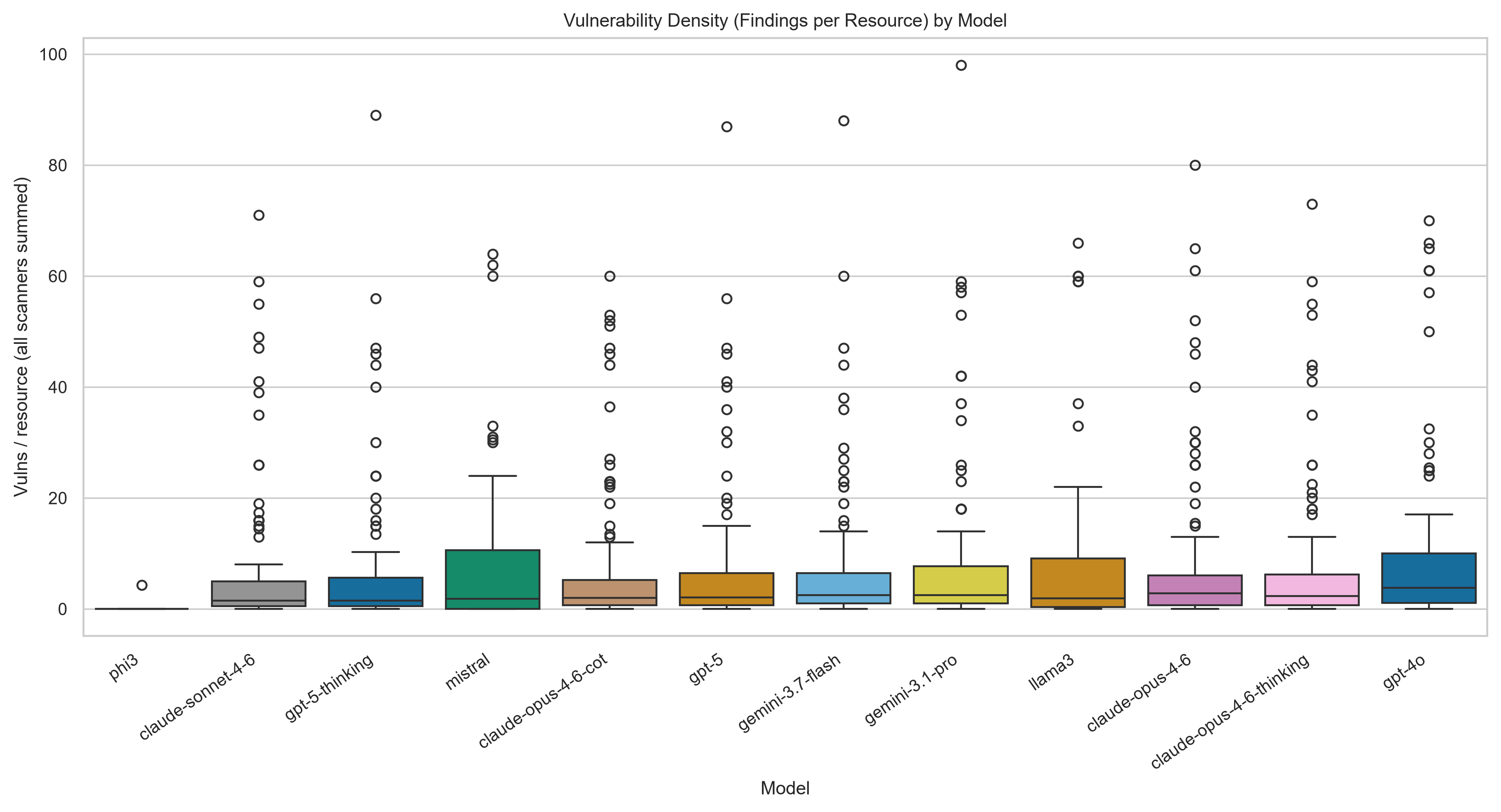}
\caption{Density distributions by configuration. The right skew motivates
rank-based inference.}
\label{fig:boxplot}
\end{figure*}

Table~\ref{tab:descriptive} reports mean and median density per configuration and
stratum. Medians are markedly lower than means throughout---the distribution is
heavily right-skewed and zero-inflated, which is why all inferential tests below
are rank-based or explicitly negative-binomial. Fig.~\ref{fig:boxplot} shows the
full distributions.

\begin{table}[t]
\caption{Vulnerability density by configuration and stratum (mean / median).
\texttt{phi3} declares almost no parseable resources in the simple stratum, so
its density is undefined there.}
\label{tab:descriptive}
\centering
\small
\begin{tabular}{@{}lcccc@{}}
\toprule
 & \multicolumn{2}{c}{Simple} & \multicolumn{2}{c}{Complex} \\
\cmidrule(lr){2-3}\cmidrule(lr){4-5}
Configuration & mean & median & mean & median \\
\midrule
claude-opus-4-6           & 12.12 & 4.00 & 4.38 & 0.68 \\
claude-opus-4-6-cot       & 11.97 & 4.00 & 1.90 & 0.65 \\
claude-opus-4-6-thinking  & 10.53 & 3.83 & 4.06 & 0.65 \\
claude-sonnet-4-6         &  8.84 & 3.37 & 5.66 & 0.49 \\
gemini-3.1-pro            & 11.24 & 4.40 & 5.55 & 1.65 \\
gemini-3.7-flash          &  9.50 & 4.00 & 4.94 & 1.17 \\
gpt-4o                    & 12.30 & 6.88 & 6.64 & 1.71 \\
gpt-5                     &  9.45 & 3.25 & 5.49 & 1.25 \\
gpt-5-thinking            &  8.44 & 3.00 & 5.65 & 1.00 \\
llama3                    & 11.29 & 4.00 & 4.61 & 1.39 \\
mistral                   & 10.98 & 5.00 & 8.49 & 5.57 \\
phi3                      & ---   & ---  & 4.33 & 4.33 \\
\midrule
\textbf{Human baseline}   & \multicolumn{4}{c}{2.66 mean / 1.00 median (n=634)} \\
\bottomrule
\end{tabular}
\end{table}

\subsection{Density must be size-matched}
\label{sec:matching}
\begin{figure}[t]
\centering
\includegraphics[width=\columnwidth]{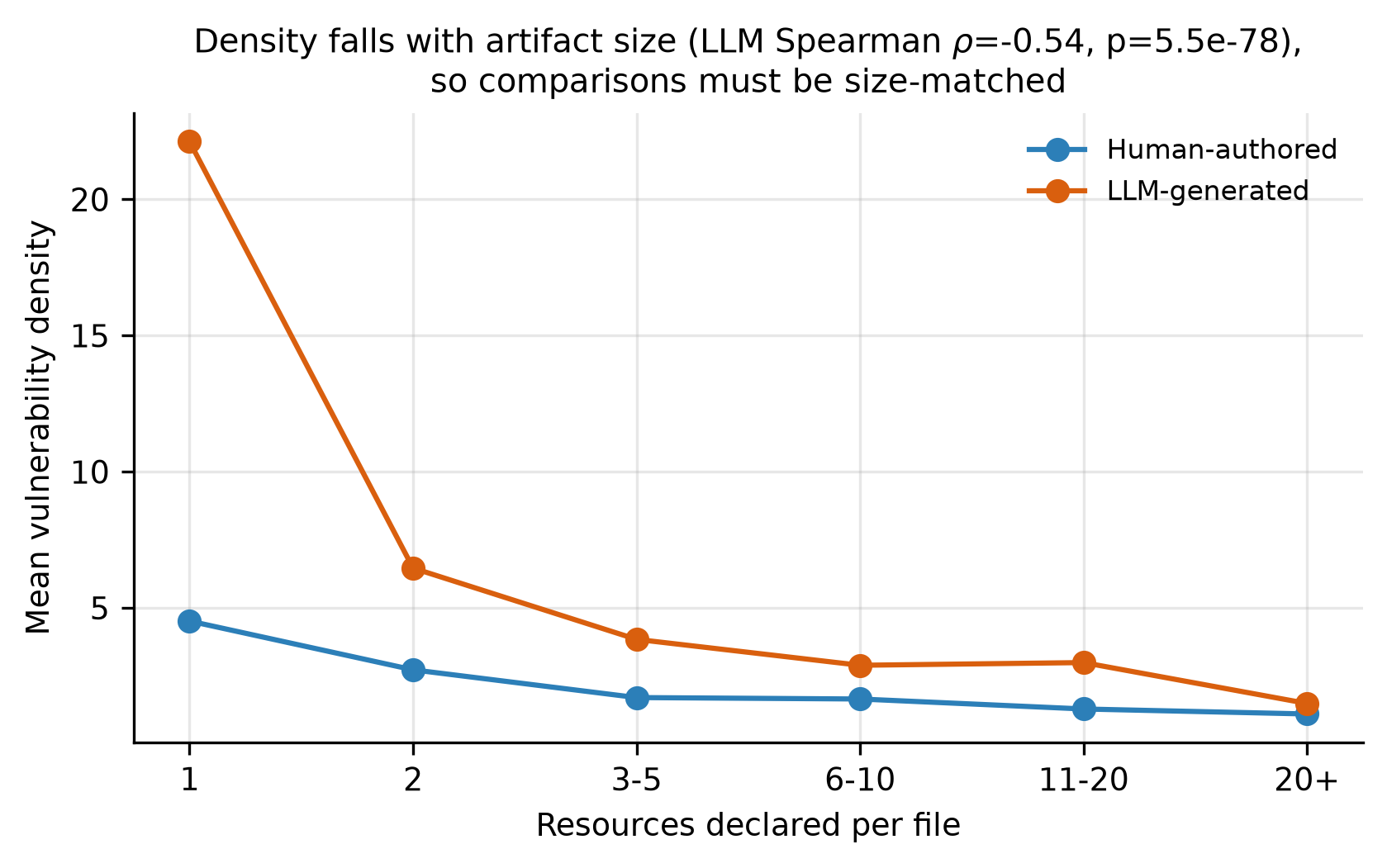}
\caption{Vulnerability density declines with artifact size in both corpora,
requiring size-matched comparison.}
\label{fig:densres}
\end{figure}

Vulnerability density falls steeply with artifact size
(Fig.~\ref{fig:densres}): in the human corpus it declines from 4.51 findings per
resource in single-resource files to 1.10 in files declaring twenty or more, and
across generated artifacts the association is strong and highly significant
(Spearman $\rho=-0.55$, $p=1.7\times10^{-78}$). Because the corpora differ in
scale---human files average 5.31 declared resources, simple generations
${\approx}3$, complex generations ${\approx}50$---any unmatched comparison
measures artifact size rather than security posture. All comparisons below are
therefore computed within resource-count strata.

\subsection{LLM-generated IaC is consistently less secure than human IaC}
\label{sec:human}
\begin{figure}[t]
\centering
\includegraphics[width=\columnwidth]{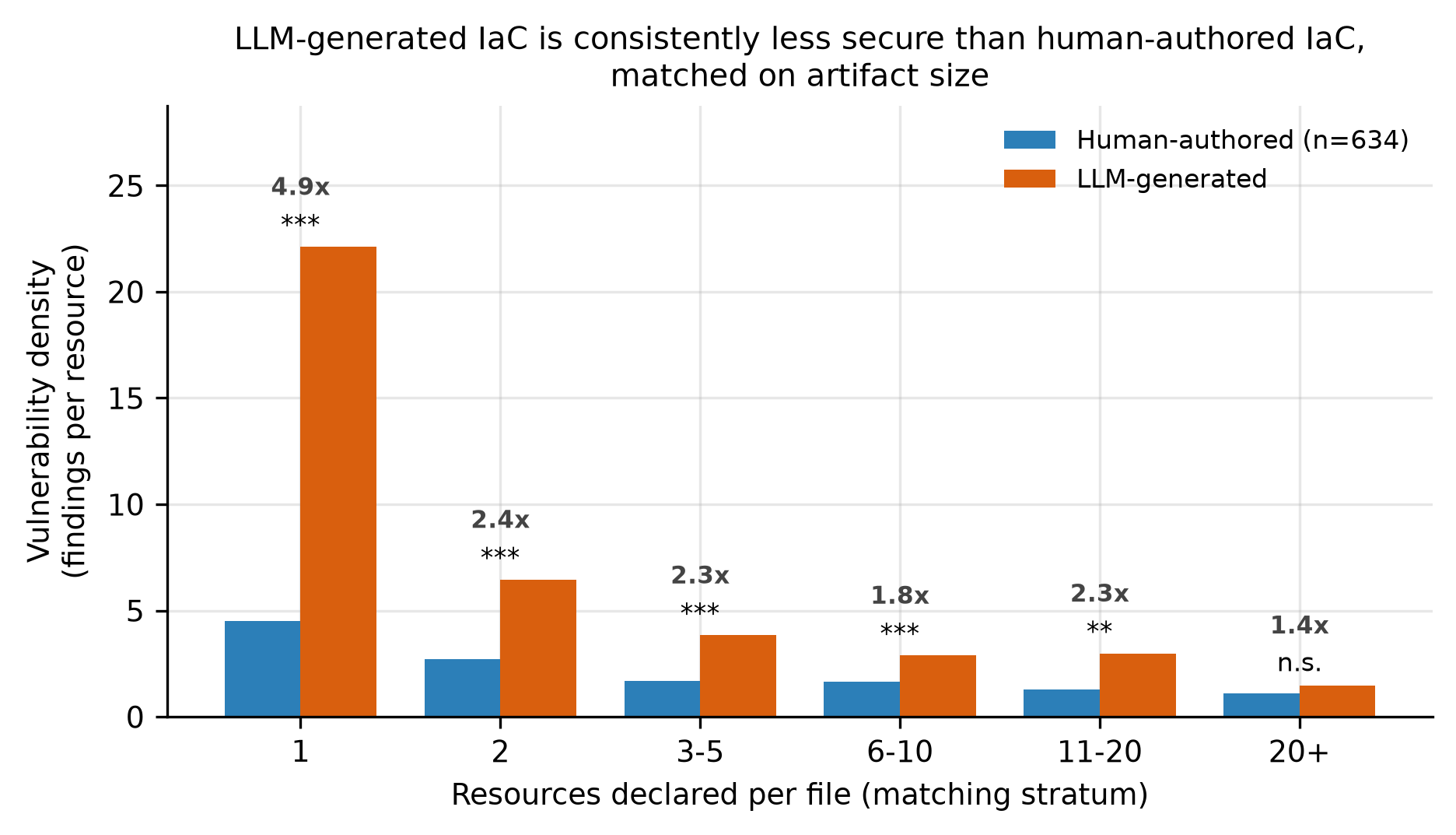}
\caption{Size-matched comparison against the human baseline.
$^{*}p<0.05$, $^{**}p<0.01$, $^{***}p<0.001$ (Mann--Whitney $U$).}
\label{fig:humanvsllm}
\end{figure}

Fig.~\ref{fig:humanvsllm} compares generated and human artifacts within matched
strata. The generated corpus exhibits significantly higher density in every
stratum except the largest, with the ratio decreasing monotonically in
artifact size: $4.9\times$ at one declared resource, $2.4\times$ at two,
$2.3\times$ at three to five, $1.8\times$ at six to ten, $2.3\times$ at eleven to
twenty, and $1.4\times$ (not significant, $p=0.058$) at twenty or more.

\begin{figure}[t]
\centering
\includegraphics[width=\columnwidth]{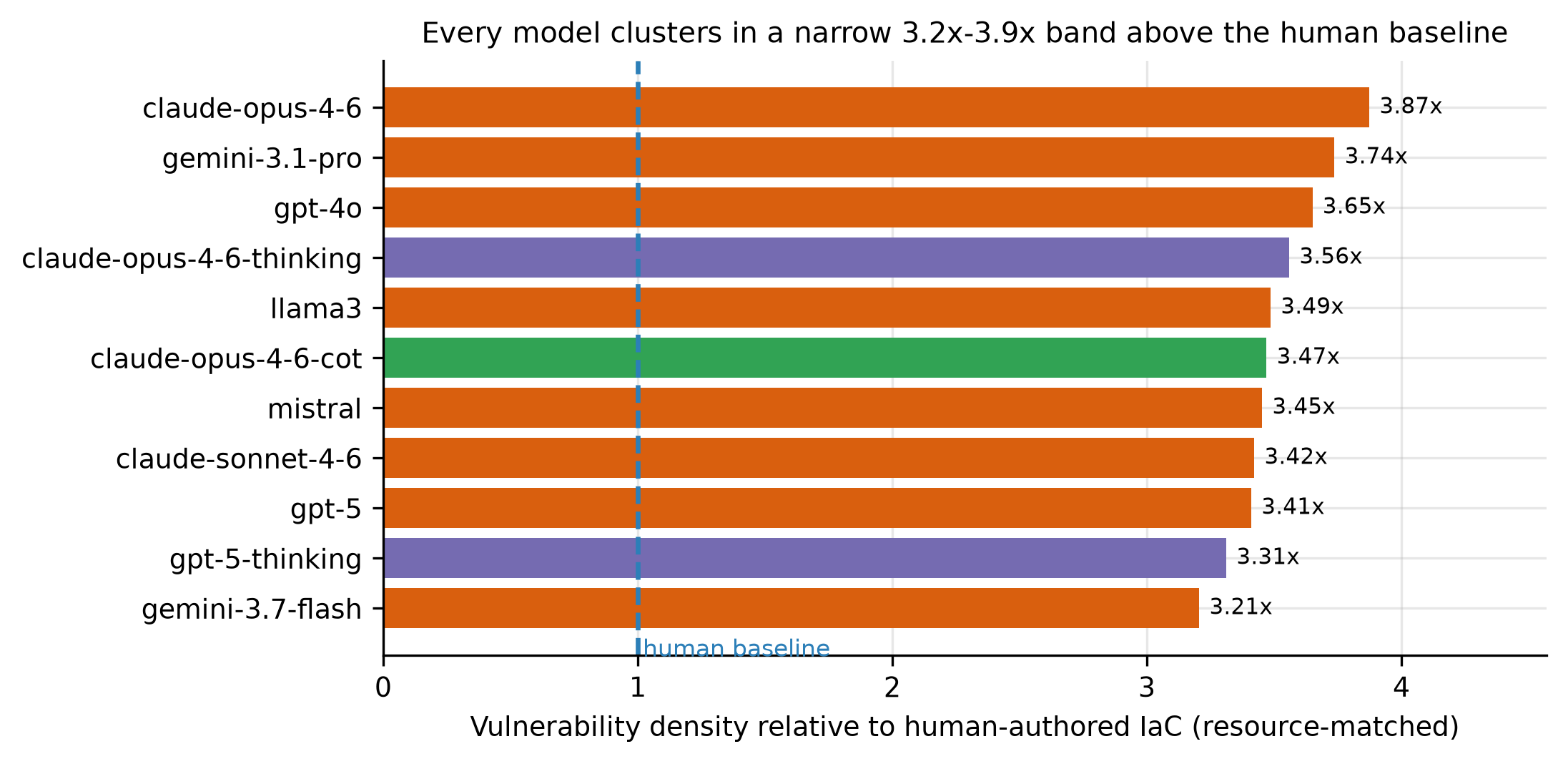}
\caption{Per-configuration density relative to the human baseline. Purple:
vendor extended thinking. Green: prompted chain-of-thought.}
\label{fig:ratio}
\end{figure}

Aggregating per configuration over shared strata (Fig.~\ref{fig:ratio}) yields a
strikingly narrow band: every configuration lies between $3.21\times$ and
$3.87\times$ the human baseline. This consistency across six vendors, open and
closed weights, and three reasoning modes suggests a property of current
LLM-generated IaC rather than of any individual model.

Notably, the gap is \emph{largest on the simplest tasks}. Where a human writes a
minimal single-resource template, models emit substantially more flagged
configuration---consistent with the structural over-generation reported in
Section~\ref{sec:structural}.

\subsection{Models differ, but the classical test cannot show it}
\label{sec:omnibus}
Skillings--Mack rejects the null of equal model performance in both strata
(simple: $\chi^2=69.3$, $df=10$, $p=6.0\times10^{-11}$, 60 blocks; complex:
$\chi^2=81.2$, $df=11$, $p=8.7\times10^{-13}$, 40 blocks).

The classical alternative fails outright. Friedman requires complete blocks, and
with 12 configurations and realistic coverage gaps \emph{no scenario has every
configuration present}: complete-case analysis retains \textbf{zero} blocks in
both strata. This is not a matter of reduced power---the standard test in this
literature~\cite{demsar2006statistical} is simply uncomputable on a design of
this shape. We regard this as a transferable methodological point: multi-model
benchmarks routinely lose cells to refusals, quota limits, and truncation, and
should default to incomplete-block statistics.

\subsection{Post-hoc pairwise comparisons}
\label{sec:posthoc}
Following Dem\v{s}ar~\cite{demsar2006statistical}, a significant omnibus result is
followed by pairwise Wilcoxon signed-rank tests~\cite{wilcoxon1945individual}
with Holm correction~\cite{holm1979simple}. Of 66 pairs per stratum, 10 survive
correction in the simple stratum and 9 in the complex stratum
(Table~\ref{tab:posthoc}). Pairwise deletion is used rather than listwise, so
each comparison uses every scenario where both configurations are present; $n$ is
reported per pair.

Two patterns are visible. In the simple stratum, \texttt{gpt-4o} and
\texttt{gpt-5-thinking} sit at opposite ends and separate from the middle of the
field. In the complex stratum, the separations are dominated by the Gemini
configurations, which differ significantly from every Claude configuration---the
strongest single result being \texttt{claude-opus-4-6-thinking} versus
\texttt{gemini-3.1-pro} ($p_{\text{adj}}=6.5\times10^{-4}$). Notably, no
comparison between the three reasoning arms of the same base model survives
correction, consistent with the modest within-model effects reported in
Section~\ref{sec:reasoning}: cross-vendor differences are larger than
reasoning-mode differences.

\begin{table}[t]
\caption{Pairwise comparisons surviving Holm correction ($\alpha=0.05$).
66 pairs tested per stratum.}
\label{tab:posthoc}
\centering
\small
\begin{tabular}{@{}llrr@{}}
\toprule
Configuration A & Configuration B & $n$ & $p_{\text{adj}}$ \\
\midrule
\multicolumn{4}{@{}l}{\emph{Simple stratum (10 of 66 significant)}} \\
claude-opus-4-6      & gpt-5-thinking   & 58 & 0.0036 \\
claude-opus-4-6      & claude-sonnet-4-6& 60 & 0.0174 \\
claude-opus-4-6-cot  & gpt-5-thinking   & 58 & 0.0026 \\
claude-opus-4-6-cot  & gpt-5            & 59 & 0.0076 \\
claude-opus-4-6-cot  & claude-sonnet-4-6& 60 & 0.0159 \\
gemini-3.1-pro       & gpt-5-thinking   & 58 & 0.0026 \\
gemini-3.7-flash     & gpt-4o           & 60 & 0.0188 \\
gpt-4o               & gpt-5-thinking   & 58 & 0.0026 \\
gpt-4o               & gpt-5            & 59 & 0.0179 \\
claude-sonnet-4-6    & gpt-4o           & 60 & 0.0448 \\
\midrule
\multicolumn{4}{@{}l}{\emph{Complex stratum (9 of 66 significant)}} \\
claude-opus-4-6-thinking & gemini-3.1-pro   & 30 & 0.00065 \\
claude-sonnet-4-6        & gpt-5            & 30 & 0.0041 \\
claude-sonnet-4-6        & gemini-3.7-flash & 32 & 0.0043 \\
claude-opus-4-6-thinking & gemini-3.7-flash & 31 & 0.0055 \\
claude-sonnet-4-6        & gemini-3.1-pro   & 31 & 0.0060 \\
claude-opus-4-6          & gemini-3.1-pro   & 32 & 0.0081 \\
claude-opus-4-6-cot      & gemini-3.1-pro   & 35 & 0.0112 \\
claude-sonnet-4-6        & gpt-5-thinking   & 29 & 0.0395 \\
claude-opus-4-6          & gemini-3.7-flash & 33 & 0.0440 \\
\bottomrule
\end{tabular}
\end{table}

\subsection{Complexity interaction}
\label{sec:interaction}
The rate model includes a $\text{configuration}\times\text{complexity}$
interaction, which tests whether a configuration's per-resource rate shifts
disproportionately between strata. Table~\ref{tab:interaction} reports the
interaction terms.

Five configurations show significant interactions, all with IRR below 1,
indicating that their simple-stratum rate is \emph{lower} than their own complex
baseline once exposure is accounted for---most strongly \texttt{gpt-5-thinking}
(IRR $0.55$, $p=4\times10^{-4}$) and \texttt{gpt-5} (IRR $0.63$,
$p=2\times10^{-4}$). The Anthropic extended-thinking arm shows essentially no
interaction (IRR $0.99$, $p=0.955$): its behaviour is stable across complexity.

We emphasise the direction here because the pre-remediation analysis reported the
opposite for one configuration---an apparent ``complexity relaxation'' effect in
which \texttt{gpt-4o} became 14$\times$ more vulnerable on simple tasks. That
estimate came from a model without an exposure offset and against an empty
reference cell (Section~\ref{sec:threats}); with both corrected, the
\texttt{gpt-4o} interaction is not significant ($p=0.275$).

\begin{table}[t]
\caption{Configuration $\times$ complexity interaction terms from the negative
binomial GEE. IRR${<}1$ indicates a lower simple-stratum rate relative to the
configuration's own complex baseline.}
\label{tab:interaction}
\centering
\small
\begin{tabular}{@{}lrr@{}}
\toprule
Configuration & IRR & $p$ \\
\midrule
gpt-5-thinking            & 0.55 & \textbf{0.00040} \\
llama3                    & 0.61 & \textbf{0.037} \\
gpt-5                     & 0.63 & \textbf{0.00024} \\
gpt-4o                    & 0.64 & 0.275 \\
mistral                   & 0.64 & 0.637 \\
gemini-3.7-flash          & 0.66 & \textbf{0.00092} \\
gemini-3.1-pro            & 0.70 & \textbf{0.000067} \\
claude-sonnet-4-6         & 0.59 & 0.164 \\
claude-opus-4-6-thinking  & 0.99 & 0.955 \\
\bottomrule
\end{tabular}
\end{table}

\subsection{Reasoning modes}
\label{sec:reasoning}
Table~\ref{tab:reasoning} reports paired within-model contrasts on the simple
stratum, where sample size is largest. Vendor extended thinking reduces density
relative to standard generation ($-13.2\%$, $p=0.012$) and, more informatively,
relative to prompted chain-of-thought ($-12.0\%$, $p=0.0013$). Prompted
chain-of-thought alone is statistically indistinguishable from standard
generation ($-1.3\%$, $p=0.238$).

\begin{table}[t]
\caption{Paired reasoning-mode contrasts. Every contrast points in the same
direction in both strata; the complex stratum is underpowered ($n\leq33$).}
\label{tab:reasoning}
\centering
\small
\begin{tabular}{@{}lcrrr@{}}
\toprule
Contrast & Stratum & $n$ & $\Delta$ density & $p$ \\
\midrule
\multirow{2}{*}{standard $\rightarrow$ ext.\ thinking}
  & simple  & 60 & $-13.2\%$ & \textbf{0.012} \\
  & complex & 31 & $-14.2\%$ & 0.176 \\
\addlinespace
\multirow{2}{*}{prompted CoT $\rightarrow$ ext.\ thinking}
  & simple  & 60 & $-12.0\%$ & \textbf{0.0013} \\
  & complex & 31 & $-16.9\%$ & 0.209 \\
\addlinespace
\multirow{2}{*}{standard $\rightarrow$ prompted CoT}
  & simple  & 60 & $-1.3\%$ & 0.238 \\
  & complex & 33 & $-4.8\%$ & 0.751 \\
\addlinespace
\multirow{2}{*}{gpt-5 $\rightarrow$ \texttt{effort=high}}
  & simple  & 58 & $-10.2\%$ & 0.153 \\
  & complex & 29 & $+2.0\%$  & 0.701 \\
\bottomrule
\end{tabular}
\end{table}

The complex stratum reproduces the direction and magnitude of every Anthropic
contrast---indeed the extended-thinking effects are slightly larger there
($-14.2\%$ and $-16.9\%$)---but with roughly half the paired observations, none
reach significance. We report these as directionally consistent but underpowered
rather than as null results; distinguishing the two would require a larger
complex stratum, which we identify as future work.

\begin{figure}[t]
\centering
\includegraphics[width=\columnwidth]{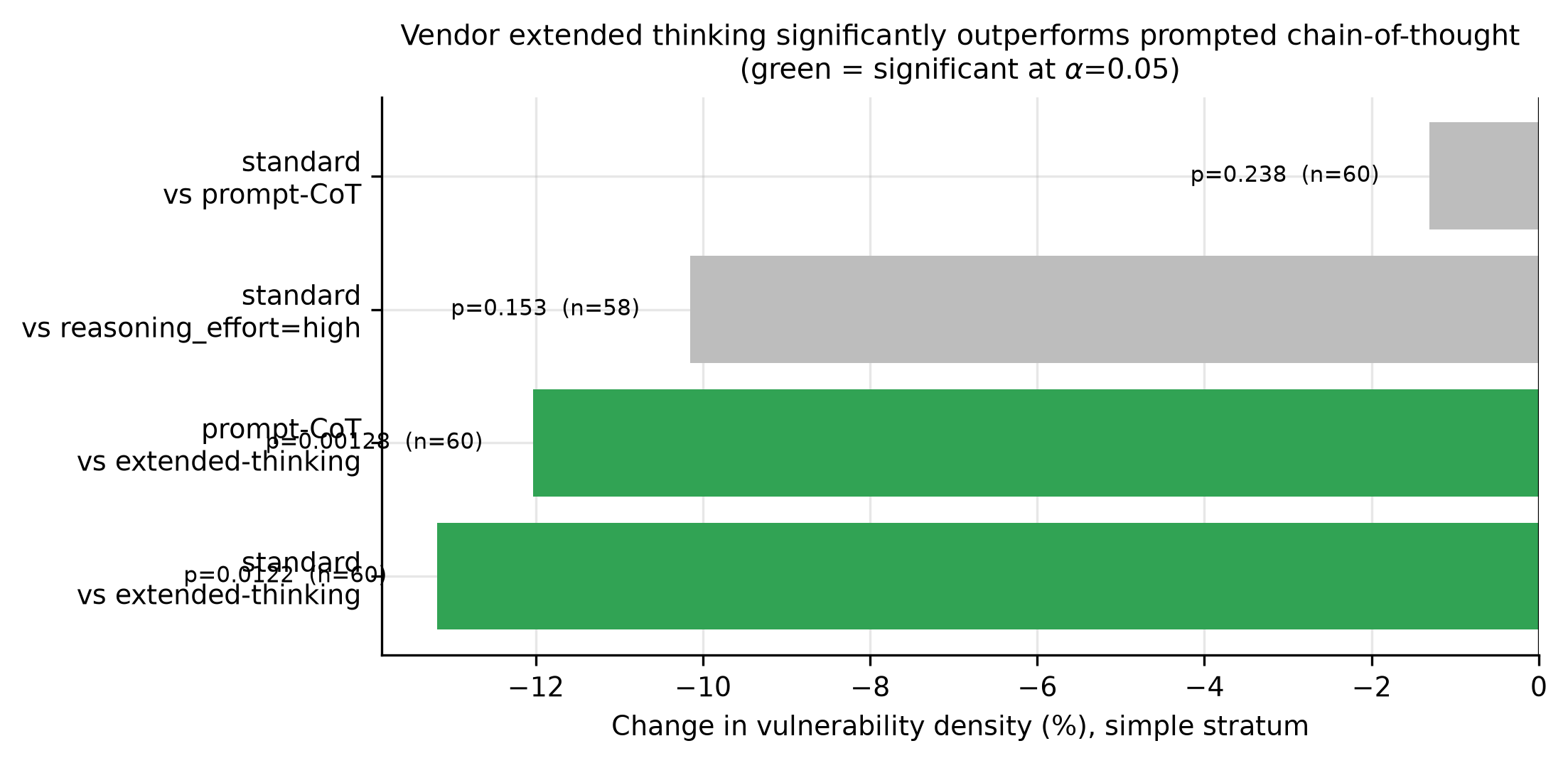}
\caption{Paired reasoning-mode contrasts on the simple stratum. Green bars are
significant at $\alpha=0.05$.}
\label{fig:contrasts}
\end{figure}

Fig.~\ref{fig:contrasts} visualizes these contrasts. The practical implication is
direct: instructing a model to ``think step by step'' does not confer the
security benefit of paying for reasoning tokens. Given that prompted
chain-of-thought is often treated as a free substitute, and that its benefits
appear concentrated in mathematical and symbolic domains~\cite{sprague2025cot},
this distinction matters for practitioner guidance.

\begin{figure}[t]
\centering
\includegraphics[width=\columnwidth]{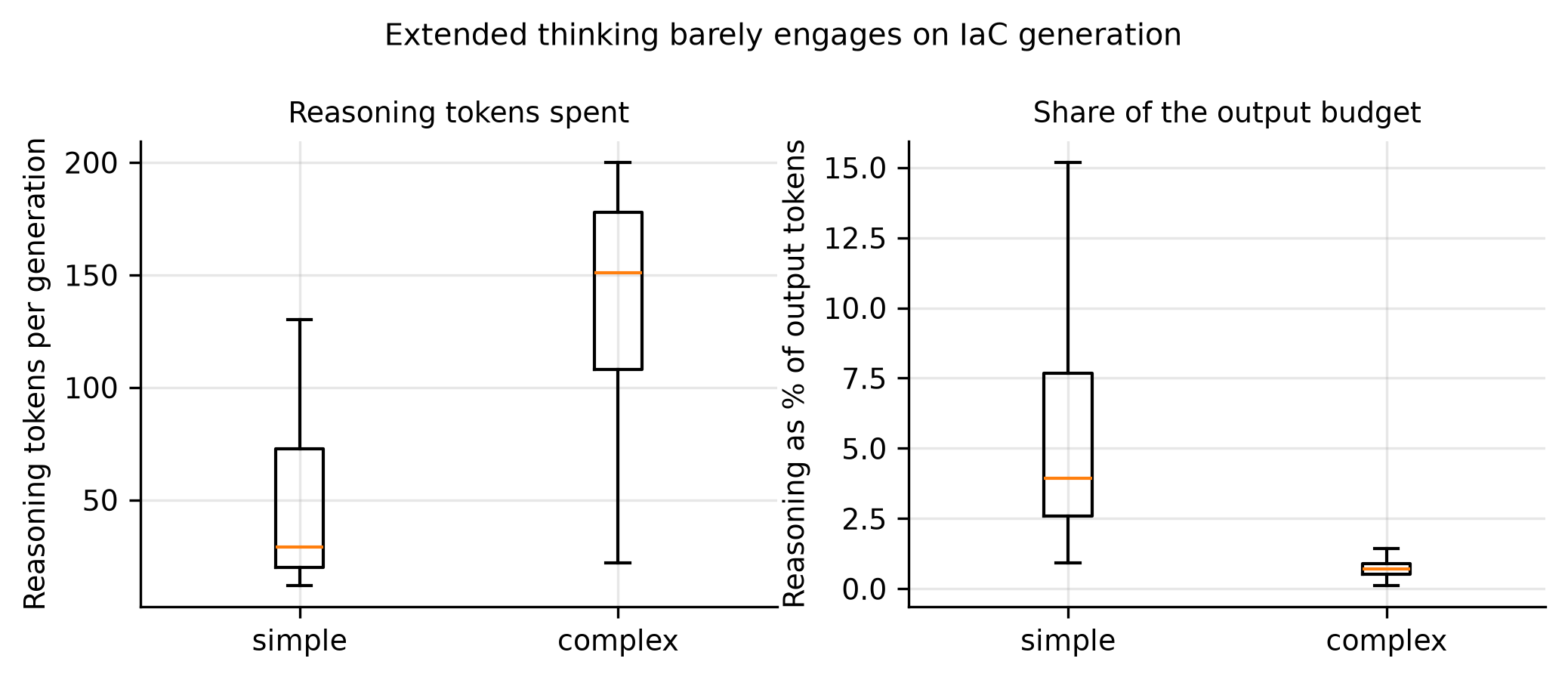}
\caption{Reasoning-token expenditure under extended thinking. The mechanism is
barely engaged on IaC generation.}
\label{fig:tokens}
\end{figure}

Instrumenting the API explains why the effect, though real, is bounded.
Fig.~\ref{fig:tokens} shows reasoning-token expenditure: a median of 29 tokens on
simple scenarios and 151 on complex ones, against median completion lengths of
886 and 18{,}533 tokens respectively---under $1\%$ of the output budget on
complex tasks, despite a generous configured allowance. The budget is a ceiling
the model may underspend, not a target. A modest measured effect should therefore
\emph{not} be read as ``reasoning does not help''; on this task class the
mechanism is barely exercised.

\subsection{Rate model}
The negative binomial GEE with exposure offset yields incidence rate ratios from
0.73 to 4.64 against a \texttt{claude-opus-4-6} reference. The prompted-CoT arm
is significantly below the reference (IRR $0.73$, 95\% CI $[0.54,0.99]$,
$p=0.041$); \texttt{phi3} is highest (IRR $4.64$, $[3.12,6.89]$,
$p=2.9\times10^{-14}$).

The \texttt{phi3} result inverts a naive reading. It records the fewest absolute
findings and appears safest by raw count, but declares almost no parseable
infrastructure; \emph{per resource} it is the worst configuration measured. This
is survivorship bias made explicit, and it is why absolute counts should not be
used for cross-model comparison.

\subsection{What models get wrong, and what the engines disagree about}
\label{sec:categories}
\begin{figure*}[t]
\centering
\includegraphics[width=0.86\textwidth]{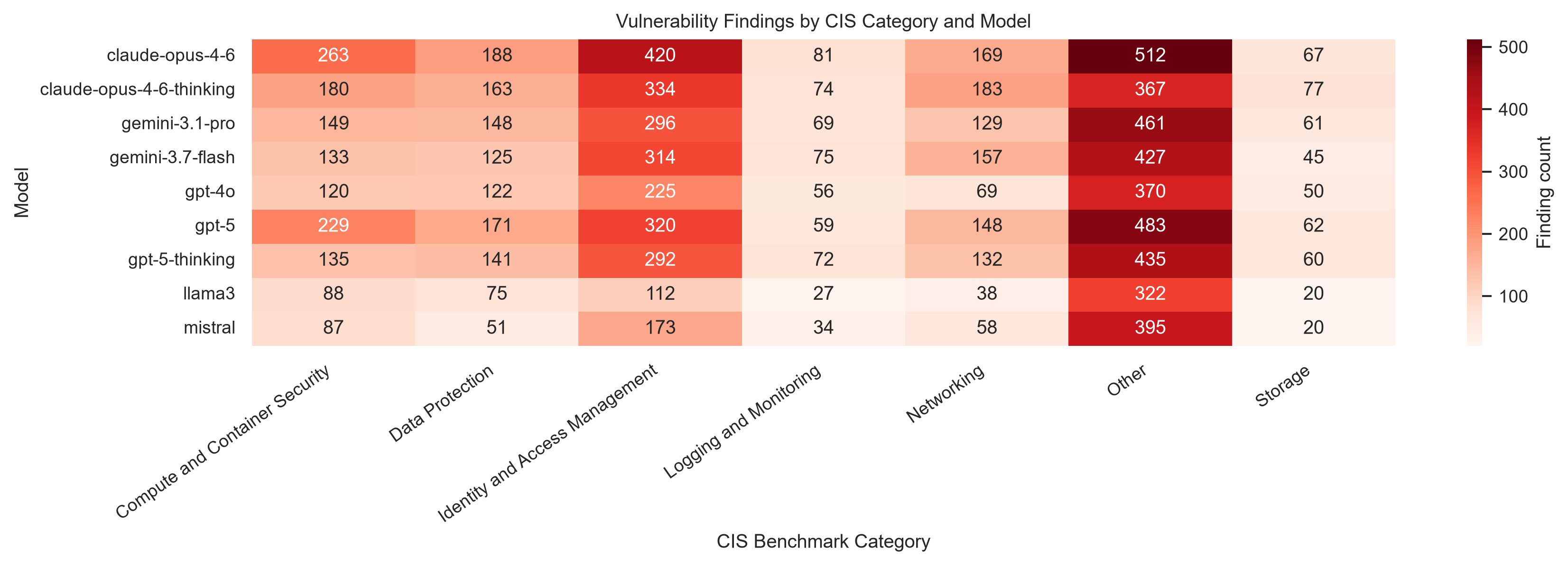}
\caption{Findings by CIS category and configuration. ``Other'' reflects rules
without a clean CIS mapping and should not be read as a vulnerability class.}
\label{fig:cis}
\end{figure*}

Mapping findings to CIS Benchmark categories~\cite{cis_benchmarks}
(Fig.~\ref{fig:cis}) shows Networking as the largest named category (4{,}554
findings), followed by Identity and Access Management (2{,}849),
Logging/Monitoring (2{,}149), and Encryption (2{,}140). A large residual
``Other'' bucket (27{,}111) is dominated by Checkov rules that do not map cleanly
onto CIS categories and is a tooling artifact rather than a vulnerability class.
The prominence of networking and IAM is consistent with the security smells
catalogued for hand-written IaC~\cite{rahman2019seven}: permissive ingress rules
and over-broad role grants remain the dominant failure modes whether the author
is human or machine.

\begin{figure}[t]
\centering
\includegraphics[width=\columnwidth]{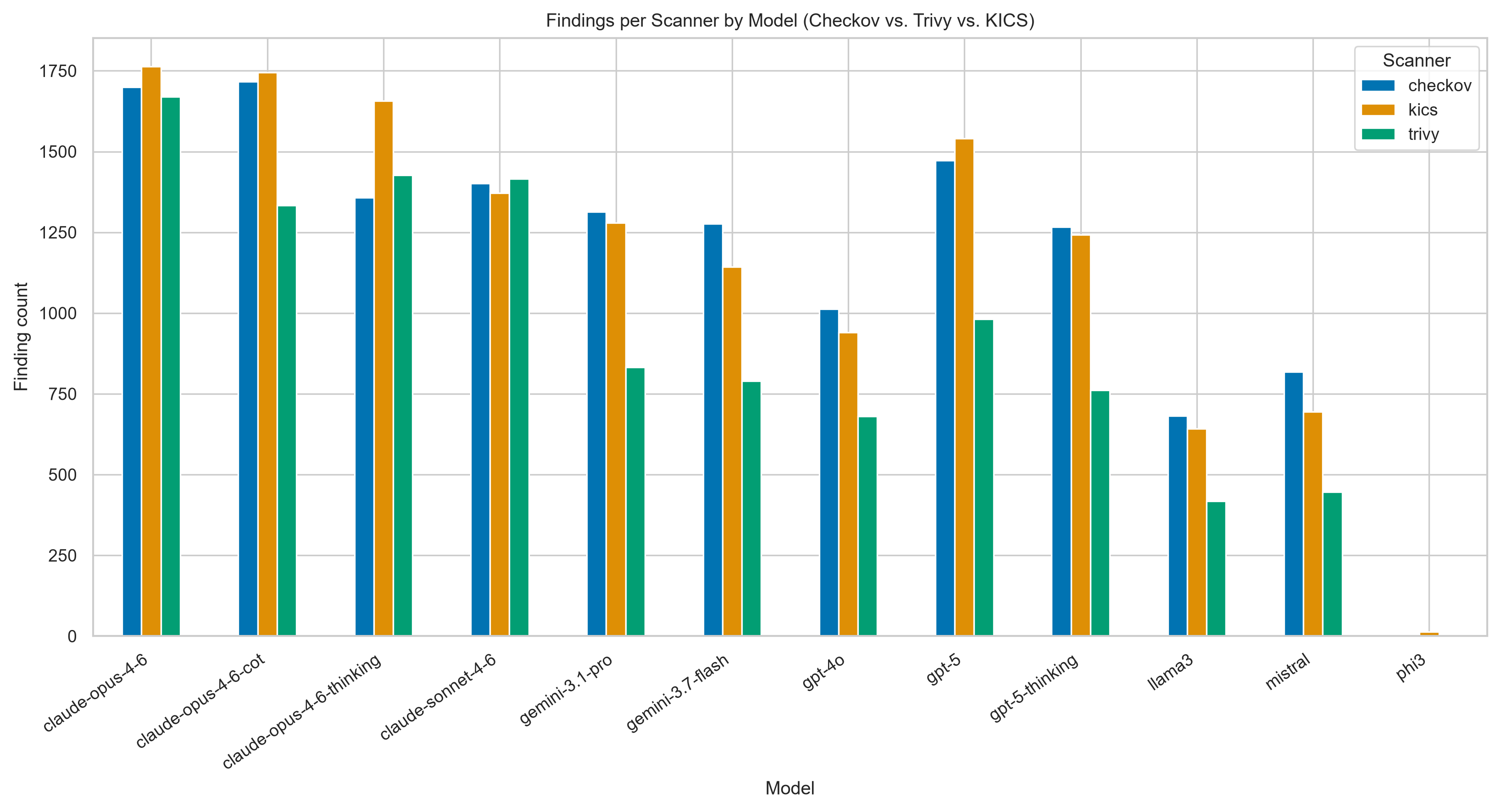}
\caption{Findings by engine. Volume disagreement across three independent
vendors motivates multi-engine evaluation.}
\label{fig:scanner}
\end{figure}

Fig.~\ref{fig:scanner} shows per-engine finding counts. The three engines
disagree substantially in volume---Checkov and KICS report comparable totals
while Trivy reports roughly 23\% fewer---which is precisely the convergent-validity
argument for using three independent vendors. A single-engine study would
inherit that engine's rule coverage as if it were ground truth. We note that our
own earlier, partially-covered corpus would have supported different conclusions
about inter-engine agreement, which is why we report coverage explicitly.

\subsection{Schema validity}
\label{sec:validity}
\begin{figure}[t]
\centering
\includegraphics[width=\columnwidth]{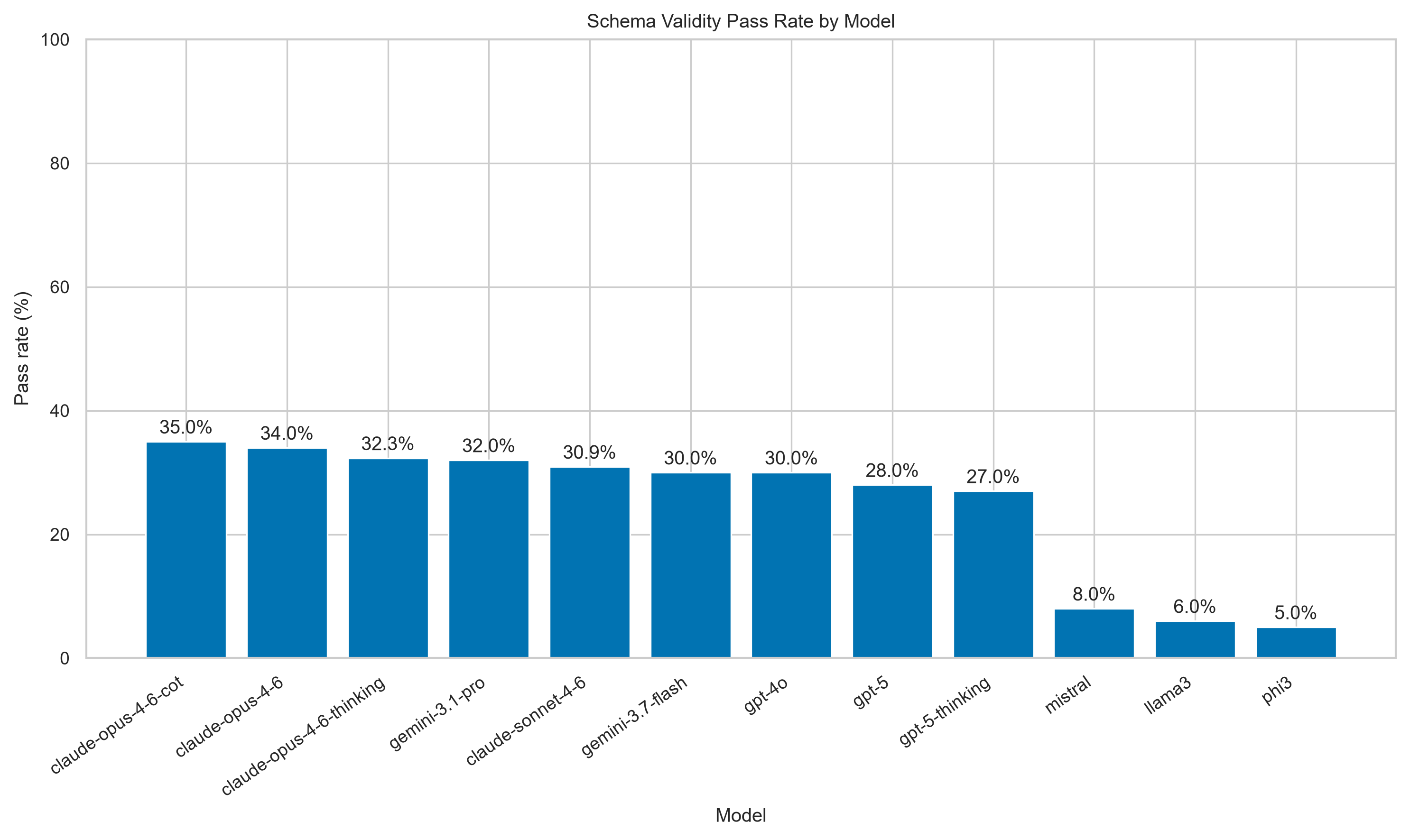}
\caption{Schema-validity pass rate by configuration.}
\label{fig:validity}
\end{figure}

Schema-validity pass rates (Fig.~\ref{fig:validity}) cluster between 27\% and
35\% for all frontier configurations and collapse for the small local models
(\texttt{mistral} 8\%, \texttt{llama3} 6\%, \texttt{phi3} 5\%). The absolute
rates are low because complex scenarios demand multi-service architectures that
must satisfy real provider schemas; they are nonetheless directly comparable
across configurations, and they are what makes the survivorship correction in
Section~\ref{sec:negative} necessary.

\subsection{Severity distribution}
\label{sec:severity}
Trivy and KICS emit CVSS-style severity tiers; Checkov does not without a
commercial subscription, so all 14{,}017 Checkov findings carry
\texttt{UNKNOWN} severity and are excluded from this analysis.
Across the remaining findings the distribution is 778 \textsc{critical},
3{,}725 \textsc{high}, 7{,}919 \textsc{medium}, 10{,}036 \textsc{low}, and
2{,}328 \textsc{info}.

Table~\ref{tab:severity} reports \textsc{critical}+\textsc{high} counts by
configuration. \texttt{claude-opus-4-6} leads (573), followed by its CoT
(503) and extended-thinking (473) arms---an ordering that mirrors the
within-model reasoning effect while remaining far above the smaller models.
\texttt{phi3} records zero severity-tiered findings, again because it emits
almost no parseable infrastructure rather than because it is secure.

These are \emph{absolute} counts and therefore confounded by output volume;
they are reported for completeness and comparability with prior work, and the
per-resource analysis in Sections~\ref{sec:human} and~\ref{sec:reasoning}
remains the basis for cross-configuration claims.

\begin{table}[t]
\caption{Severity-tiered findings (Trivy + KICS only).}
\label{tab:severity}
\centering
\small
\begin{tabular}{@{}lrrrrr@{}}
\toprule
Configuration & Crit. & High & Med. & Low & C+H \\
\midrule
claude-opus-4-6          & 124 & 449 & 971 & 1517 & \textbf{573} \\
claude-opus-4-6-cot      &  92 & 411 & 908 & 1319 & 503 \\
gpt-5                    &  84 & 400 & 800 &  935 & 484 \\
claude-opus-4-6-thinking &  92 & 381 & 826 & 1432 & 473 \\
gemini-3.7-flash         &  78 & 342 & 723 &  674 & 420 \\
claude-sonnet-4-6        &  83 & 330 & 865 & 1400 & 413 \\
gemini-3.1-pro           &  66 & 325 & 745 &  747 & 391 \\
gpt-5-thinking           &  65 & 312 & 775 &  707 & 377 \\
gpt-4o                   &  46 & 327 & 556 &  519 & 373 \\
mistral                  &  28 & 280 & 377 &  367 & 308 \\
llama3                   &  20 & 168 & 372 &  419 & 188 \\
phi3                     &   0 &   0 &   1 &    0 & 0 \\
\bottomrule
\end{tabular}
\end{table}

\begin{figure}[t]
\centering
\includegraphics[width=\columnwidth]{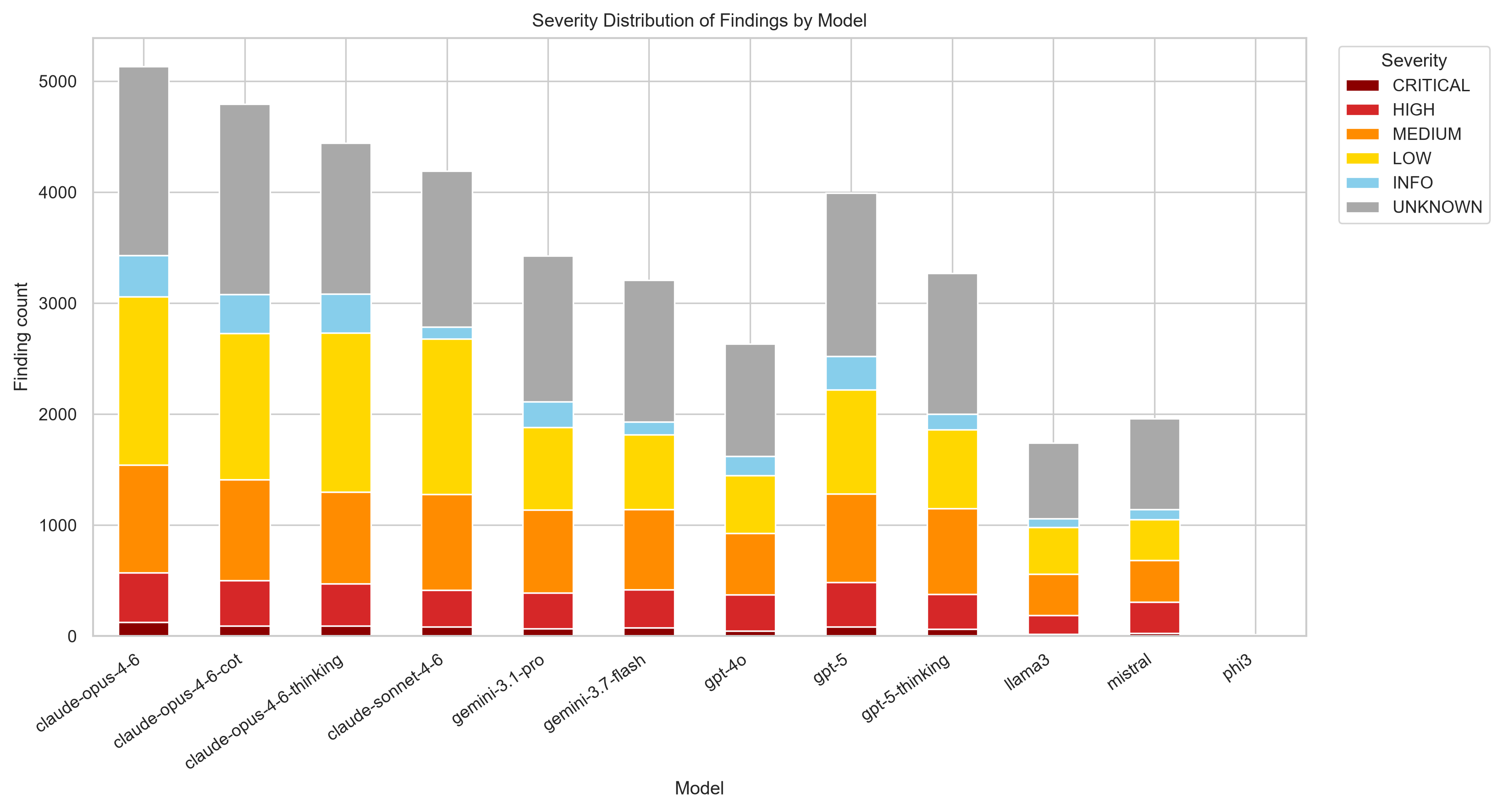}
\caption{Severity distribution by configuration (Trivy + KICS).}
\label{fig:severity}
\end{figure}

\subsection{IaC format}
\label{sec:format}
Table~\ref{tab:format} reports density by target format. We report these figures
with an explicit caveat: the resource-counting parser is format-dependent, and
its mean counts differ sharply across formats (Terraform 23.17, ARM 2.31,
Kubernetes 1.15, CloudFormation 0.90). Because resource count is the
\emph{denominator}, cross-format density values are \textbf{not} comparable, and
the large apparent densities for Kubernetes and CloudFormation reflect
under-counted denominators rather than dramatically worse security.

The Terraform subset---836 of 1{,}196 artifacts, and the format where the parser
is most reliable---is the sound basis for cross-format inference, and the human
corpus is likewise dominated by Terraform and CloudFormation. Within the human
corpus, per-format densities are stable (2.62--2.75 for the three well-populated
formats), which suggests the human baseline is not itself driven by format mix.
Establishing per-format model comparisons on equal footing requires a
format-normalized resource counter, which we leave to future work.

\begin{table}[t]
\caption{Density by IaC format. Cross-format values are NOT comparable: the
resource-count denominator is parser- and format-dependent.}
\label{tab:format}
\centering
\small
\begin{tabular}{@{}lrrr@{}}
\toprule
Format & $n$ & mean resources & density \\
\midrule
Terraform      & 836 & 23.17 & 4.60 \\
ARM            & 120 &  2.31 & 3.23 \\
CloudFormation & 120 &  0.90 & 18.66$^\dagger$ \\
Kubernetes     & 120 &  1.15 & 29.11$^\dagger$ \\
\midrule
\multicolumn{4}{@{}l}{\emph{Human corpus}} \\
\texttt{.tf}   & 235 &  2.29 & 2.75 \\
\texttt{.json} & 188 &  7.06 & 2.65 \\
\texttt{.yaml} & 198 &  7.18 & 2.62 \\
\bottomrule
\multicolumn{4}{@{}l}{\footnotesize $^\dagger$ inflated by an under-counted denominator.}
\end{tabular}
\end{table}

\subsection{Structural divergence}
\label{sec:structural}
Two-sample KS tests reject distributional equality with human-authored IaC for
every configuration on every structural metric (36/36 tests, all $p<0.05$;
Table~\ref{tab:ks}). Generated templates declare 13--36 resources on average
against 5.31 for humans. \texttt{gpt-4o} is closest to human structure across all
three metrics ($D=0.160$--$0.189$), and is the only configuration whose
resource-count divergence is marginal rather than overwhelming ($p=0.021$ against
$p<10^{-4}$ elsewhere). \texttt{phi3} is most divergent ($D=0.990$), reflecting
near-total generation failure rather than a stylistic difference.

Models systematically over-generate infrastructure relative to human engineers.
This matters for the central result: because density normalizes by resource
count, over-generation alone does not explain the security gap in
Section~\ref{sec:human}---models emit both more infrastructure \emph{and} more
findings per unit of infrastructure.

\begin{figure}[t]
\centering
\includegraphics[width=\columnwidth]{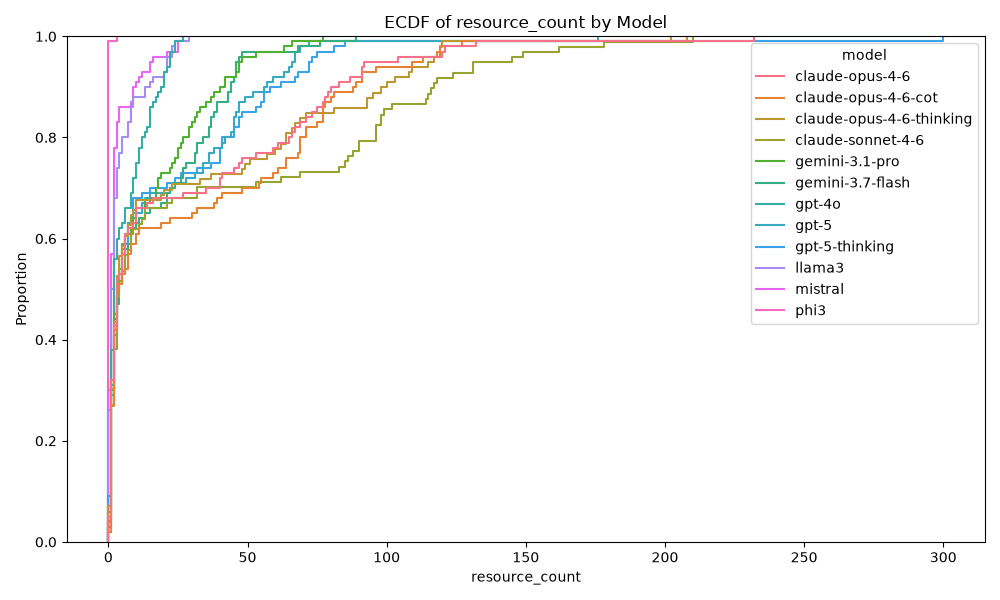}
\caption{Empirical CDFs of declared-resource count. Generated distributions sit
to the right of the human corpus throughout.}
\label{fig:ecdf}
\end{figure}

\begin{table}[t]
\caption{Kolmogorov--Smirnov $D$ against the human corpus. All 36 tests reject at
$p<0.05$.}
\label{tab:ks}
\centering
\small
\begin{tabular}{@{}lrrr@{}}
\toprule
Configuration & AST depth & Resource count & Diversity \\
\midrule
gpt-4o                   & 0.189 & \textbf{0.160} & \textbf{0.161} \\
llama3                   & 0.190 & 0.260 & 0.254 \\
gemini-3.1-pro           & 0.219 & 0.246 & 0.231 \\
gpt-5-thinking           & 0.229 & 0.251 & 0.234 \\
claude-opus-4-6          & 0.259 & 0.292 & 0.268 \\
gemini-3.7-flash         & 0.259 & 0.275 & 0.261 \\
gpt-5                    & 0.259 & 0.265 & 0.247 \\
claude-opus-4-6-thinking & 0.273 & 0.274 & 0.264 \\
mistral                  & 0.295 & 0.300 & 0.294 \\
claude-opus-4-6-cot      & 0.309 & 0.335 & 0.308 \\
claude-sonnet-4-6        & 0.322 & 0.301 & 0.306 \\
phi3                     & 0.990 & 0.990 & 0.984 \\
\bottomrule
\end{tabular}
\end{table}

\subsection{Human evaluation and judge calibration}
\label{sec:human-eval}
Three cloud-security practitioners independently scored a stratified sample of
scenarios on four criteria. Inter-rater agreement (Fleiss'
$\kappa$~\cite{fleiss1971measuring}) was fair on real-world plausibility
($0.391$) and hallucination flagging ($0.266$), slight on architectural
coherence ($0.210$), and negligible on security-test relevance ($0.059$).
Under the Landis--Koch scale~\cite{landis1977measurement} the last is barely
above chance: experienced engineers do not agree on whether a scenario poses a
meaningful security decision. This is a substantive result about the difficulty
of human security evaluation, not merely a limitation of our panel.

Against the human consensus, the LLM judge is substantial on hallucination
detection (94.4\% exact, Cohen's $\kappa=0.640$~\cite{cohen1960coefficient}) and
near-perfect on plausibility (quadratic-weighted $\kappa=0.795$, 100\% within one
point), but near chance on architectural coherence ($\kappa=0.177$, 27.8\%
exact). The boundary is actionable: automated judges are usable for factual
verification, not for architectural assessment.

\subsection{Two negative results}
\label{sec:negative}
First, the intuitive \emph{validity--security} hypothesis---that models better
at producing deployable code produce more vulnerable code---is unsupported.
Across configurations, schema-validity pass rate and vulnerability density are
uncorrelated ($r=0.158$, $p=0.625$; Spearman $\rho=0.098$, $p=0.761$). Pass rates
cluster between 27\% and 35\% for all frontier configurations with no
accompanying density relationship. Only the survivorship component holds:
\texttt{phi3} passes 5\% and produces almost no parseable infrastructure.

Second, as reported in Section~\ref{sec:omnibus}, complete-case Friedman is
uncomputable on this design.

\subsection{Complexity stratification is measurable in the artifacts}
\label{sec:stratacheck}
The simple/complex split is a design choice and should be validated rather than
assumed. KS tests comparing each configuration's simple-stratum output against
its own complex-stratum output reject equality for the structural metrics across
configurations, confirming that the two tiers elicit materially different
architectures rather than differing only in prompt wording. Separately, KS tests
of each configuration against the pooled remainder find 30 of 48 tests
significant at $\alpha=0.05$, indicating that configurations occupy
distinguishable structural niches. \texttt{gpt-5} and \texttt{gpt-5-thinking} are
the least distinguishable from the field ($D=0.134$, $p=0.068$)---unsurprising,
since they share a base model---while \texttt{phi3} is essentially disjoint
($D=0.901$).

\subsection{Summary of statistical procedures}
\label{sec:testsummary}
Table~\ref{tab:alltests} consolidates every inferential procedure reported in
this paper: what each measures, why that test was chosen over alternatives, and
what it returned. We include the tests that produced null and negative results,
and the one test that could not be computed at all, since selectively reporting
only significant procedures is itself a source of bias.

\begin{table*}[!tp]
\caption{All statistical procedures performed, what each measures, and its
outcome. ``$\rightarrow$'' gives the interpretation. Tests are grouped by the
question they answer; null and negative results are included.}
\label{tab:alltests}
\centering
\scriptsize
\renewcommand{\arraystretch}{1.08}
\begin{tabular}{@{}p{2.7cm}p{3.5cm}p{4.0cm}p{6.4cm}@{}}
\toprule
\textbf{Test} & \textbf{What it measures} & \textbf{Why this test} & \textbf{Result and outcome} \\
\midrule
\multicolumn{4}{@{}l}{\textbf{\emph{Q1. Are LLMs less secure than human engineers?}}} \\
\addlinespace[2pt]
Spearman rank correlation
& Association between vulnerability density and artifact size
& Non-parametric; densities are skewed and the relationship need not be linear
& $\rho=-0.546$, $p=1.7\times10^{-78}$ $\rightarrow$ density falls sharply with size, so \emph{unmatched} comparisons measure artifact size, not security. Motivates all matching below. \\
Mann--Whitney $U$ (per size stratum)
& LLM vs.\ human density within matched resource-count bins
& Two independent samples, non-normal; matching removes the size confound
& Significant in 5 of 6 bins ($p<0.05$); ratio $4.9\times$ (1 resource) $\rightarrow$ $1.4\times$ (20+, n.s.) $\rightarrow$ \textbf{gap is real and largest on the simplest artifacts}. \\
Aggregate matched ratio
& Per-configuration density relative to human, weighted across bins
& Summarizes the stratified comparison in one figure per configuration
& All 12 configurations in $3.21\times$--$3.87\times$ $\rightarrow$ \textbf{consistent across vendors, weights, and reasoning modes}. \\
\addlinespace[3pt]
\multicolumn{4}{@{}l}{\textbf{\emph{Q2. Do the configurations differ from each other?}}} \\
\addlinespace[2pt]
Skillings--Mack
& Omnibus equality of configurations across scenarios (blocked design)
& Generalizes Friedman to \emph{incomplete} blocks; required here because no scenario has all 12 configurations present
& Simple: $\chi^2=69.3$, $df=10$, $p=6.0\times10^{-11}$ (60 blocks). Complex: $\chi^2=81.2$, $df=11$, $p=8.7\times10^{-13}$ (40 blocks) $\rightarrow$ \textbf{configurations differ}. \\
Friedman (complete-case)
& Same hypothesis, listwise deletion
& Classical alternative; reported to justify its replacement
& \textbf{Uncomputable}: 0 usable blocks in both strata $\rightarrow$ negative methodological result (Section~\ref{sec:negative}). \\
Wilcoxon signed-rank + Holm
& Pairwise configuration differences after a significant omnibus
& Paired (same scenarios); Holm controls family-wise error over 66 pairs
& 10/66 significant (simple), 9/66 (complex) $\rightarrow$ cross-vendor gaps dominate; \textbf{no same-model reasoning-arm pair survives correction}. \\
\addlinespace[3pt]
\multicolumn{4}{@{}l}{\textbf{\emph{Q3. Does reasoning improve security?}}} \\
\addlinespace[2pt]
Wilcoxon signed-rank (paired contrasts)
& Within-model effect of each reasoning mode on density
& Same model, same scenarios, one variable toggled; non-normal paired data
& Extended thinking $-13.2\%$ ($p=0.012$); CoT$\rightarrow$thinking $-12.0\%$ ($p=0.0013$); prompted CoT alone $-1.3\%$ (n.s.) $\rightarrow$ \textbf{vendor reasoning helps modestly; prompted CoT does not}. \\
Reasoning-token share
& Fraction of output budget spent on reasoning
& Direct instrumentation rather than inference
& ${<}1\%$ of output tokens $\rightarrow$ explains why the effect is bounded. \\
\addlinespace[3pt]
\multicolumn{4}{@{}l}{\textbf{\emph{Q4. What drives the rate, controlling for exposure?}}} \\
\addlinespace[2pt]
Negative binomial GEE ($\log$ exposure offset)
& Per-resource finding rate by configuration and complexity
& Counts are overdispersed; offset makes coefficients \emph{rates} not volumes; GEE handles repeated scenarios
& Converged with exchangeable correlation $\rightarrow$ IRRs interpretable per resource. \\
Overdispersion check
& Variance-to-mean ratio of the count outcome
& Determines whether Poisson is admissible
& $\approx\!130\times$ $\rightarrow$ Poisson rejected; NB2 required (Section~\ref{sec:threats}). \\
Configuration $\times$ complexity interaction
& Whether a configuration's rate shifts disproportionately between strata
& Tests differential degradation directly
& 5 of 11 significant, all IRR${<}1$ $\rightarrow$ lower simple-stratum rates once exposure is controlled; the previously reported ``complexity relaxation'' effect does \textbf{not} survive ($p=0.275$). \\
\addlinespace[3pt]
\multicolumn{4}{@{}l}{\textbf{\emph{Q5. Is generated IaC structurally like human IaC?}}} \\
\addlinespace[2pt]
Two-sample Kolmogorov--Smirnov (vs.\ human)
& Distributional equality on AST depth, resource count, diversity
& Distribution-free; compares whole distributions, not just means
& 36/36 tests reject ($p<0.05$) $\rightarrow$ \textbf{all configurations structurally diverge}; \texttt{gpt-4o} closest ($D=0.160$), \texttt{phi3} most divergent ($D=0.990$). \\
KS (configuration vs.\ field)
& Whether each configuration occupies a distinct structural niche
& Same rationale, applied pairwise against the pooled remainder
& 30/48 significant $\rightarrow$ configurations are structurally distinguishable. \\
KS (simple vs.\ complex)
& Whether the complexity strata actually differ in the artifacts
& Validates the stratification rather than assuming it
& Rejects equality $\rightarrow$ \textbf{stratification is real}, not just prompt wording. \\
\addlinespace[3pt]
\multicolumn{4}{@{}l}{\textbf{\emph{Q6. Is the evaluation itself trustworthy?}}} \\
\addlinespace[2pt]
Fleiss' $\kappa$
& Agreement among three human raters per criterion
& Chance-corrected agreement for ${>}2$ raters
& Plausibility $0.391$; hallucination $0.266$; coherence $0.210$; \textbf{security relevance $0.059$ (near chance)} $\rightarrow$ some criteria are not reliably measurable by humans. \\
Cohen's $\kappa$ (binary)
& Human consensus vs.\ LLM judge on the hallucination flag
& Chance-corrected agreement for a binary label
& $\kappa=0.640$, 94.4\% exact $\rightarrow$ \textbf{judge is substantially reliable for factual verification}. \\
Quadratic-weighted $\kappa$ (ordinal)
& Human consensus vs.\ LLM judge on the 1--5 criteria
& Penalizes large ordinal disagreements more than adjacent ones
& Plausibility $0.795$; relevance $0.489$; \textbf{coherence $0.177$} $\rightarrow$ judge is unreliable for architectural judgment. \\
\addlinespace[3pt]
\multicolumn{4}{@{}l}{\textbf{\emph{Q7. Negative result: the validity--security paradox}}} \\
\addlinespace[2pt]
Pearson correlation
& Schema-validity pass rate vs.\ vulnerability density across configurations
& Tests the prior hypothesis that more deployable models are more vulnerable
& $r=0.158$, $p=0.625$ $\rightarrow$ \textbf{hypothesis unsupported}; reported as a negative result. \\
\bottomrule
\end{tabular}
\end{table*}

\section{Discussion}
\label{sec:discussion}

\subsection{What the human anchor changes}
The practical question facing an engineering team is not whether generated IaC
contains findings---all IaC does---but whether adopting a model changes their
risk relative to writing it themselves. Model-only benchmarks cannot answer this.
Our size-matched comparison gives a concrete figure: roughly a $3.5\times$
increase in findings per declared resource, consistent across every vendor
tested.

The consistency is the surprising part. Configurations spanning six vendors,
open and closed weights, three reasoning modes, and a $10\times$ range in
schema-validity pass rate nonetheless fall within a $3.21$--$3.87\times$ band.
This argues against a model-selection remedy: choosing a different frontier model
does not meaningfully change the security profile of the generated
infrastructure. It also argues that the gap originates upstream of any particular
vendor's alignment work---plausibly in the training distribution, which is
dominated by public repository code where permissive defaults are
common~\cite{rahman2019seven}.

\subsection{The gap is largest where review is weakest}
The ratio declines monotonically with artifact size, from $4.9\times$ on
single-resource templates to a non-significant $1.4\times$ on templates declaring
twenty or more resources. This inverts the intuition that complex generations are
the risky ones.

The practical implication is uncomfortable. Small generated snippets are exactly
the artifacts least likely to receive careful review---a five-line storage bucket
definition reads as obviously correct---yet they carry the largest relative
security penalty. Large generated architectures attract scrutiny and, per
resource, are closer to human-authored code. Review effort is therefore
misallocated by default: it scales with artifact size, while relative risk scales
inversely.

\subsection{Reasoning modes are not a solution, but they are not nothing}
Extended thinking produces a real, statistically significant reduction
($-13.2\%$), significantly outperforms the prompt-based substitute
($-12.0\%$, $p=0.0013$), and moves no configuration near the human baseline. A
$13\%$ reduction against a $3.5\times$ deficit does not close the gap.

The token instrumentation suggests why, and also where the ceiling might not be.
Because the model spends under $1\%$ of its output budget on reasoning for this
task class, the observed effect is produced by a barely-engaged mechanism. Two
readings are consistent with our data: either IaC generation does not elicit
reasoning because it is perceived as a recall task rather than a problem-solving
one, or current reasoning training does not transfer to declarative
configuration. Distinguishing these requires interventions that force higher
reasoning expenditure, which is a natural follow-up.

For practitioners the immediate guidance is narrower and firmer: if a reasoning
mode is available, enabling it is measurably better than instructing the model to
``think step by step,'' and the latter provides no measurable security benefit at
all.

\subsection{Absolute counts mislead}
Three results in this study point the same way. \texttt{phi3} records the fewest
absolute findings and the highest per-resource rate. Severity-tiered counts rank
the most capable configuration worst, purely because it emits the most code.
Cross-format densities are dominated by a parser-dependent denominator. Each is a
case where the intuitive metric inverts the correct one.

We therefore recommend that IaC security benchmarks report exposure-normalized
rates with an explicit denominator definition, and treat absolute counts as
descriptive only. The negative binomial model with a $\log(\textit{resources})$
offset is the form we found necessary: overdispersion of $130\times$ rules out
Poisson, and omitting the offset converts a rate comparison into a code-volume
comparison.

\subsection{Benchmark descriptions drift from benchmark data}
\label{sec:drift}
The prompt-prescriptiveness error in this paper's first preprint is worth
dwelling on, because its failure mode generalizes and is not specific to us.

Every \emph{numeric} claim in that preprint was correct. Corpus counts, severity
totals, scanner coverage, and per-configuration densities all reproduced exactly
under external audit, because each was generated programmatically from the
released artifacts rather than transcribed by hand. The errors were in
\emph{prose}: a sentence asserting that prompts specify function only, and a
count of vendors stated in words. Neither is a number in a table, so neither was
covered by any check we had, and both survived internal review precisely because
they described the data rather than being computed from it.

This asymmetry seems general. Benchmark papers increasingly ship reproducible
pipelines, and reviewers increasingly ask whether numbers regenerate. Far less
attention goes to whether the paper's \emph{characterization} of its own data is
accurate: what the prompts ask for, how the corpus was selected, what the
conditions actually differ in. Those claims carry most of a benchmark's
interpretive weight -- they determine what the numbers are evidence \emph{of} --
and they are the least mechanically verified part of the artifact.

Our response is to treat prose claims as testable. The released pipeline includes
a verification pass that parses assertions from the manuscript and checks them
against the artifacts, failing the build on disagreement: whether the stated
vendor count matches the model registry, whether the prompt-class distribution
matches the released classification, and whether the paper reasserts the specific
claim it previously got wrong. Both historical errors are caught by it. The
approach is crude, and we do not suggest it generalizes to every claim a paper
makes, but the class of claim it covers -- statements about the composition of a
released dataset -- is exactly the class that went wrong here.

We also note the mechanism that surfaced this: an unsolicited external audit of
released artifacts, by a researcher with no prior connection to the
work~\cite{iacguardv}. That the audit was possible at all is a function of
releasing prompts, generations, raw scanner output, and analysis code together.
A benchmark that released only aggregate tables would have kept this error
indefinitely, since nothing in the tables was wrong. Artifact release is usually
justified on reproducibility grounds; this episode suggests it also functions as
error correction, and that the value is realized only when the release is
complete enough for a third party to re-derive the interpretation rather than
merely re-run the computation.

\subsection{Implications for evaluation methodology}
Two methodological results generalize beyond IaC. First, complete-case Friedman
is uncomputable on a realistic multi-model design: with twelve configurations,
no scenario had complete coverage. Benchmarks that lose cells to refusals, quota
exhaustion, or truncation---which is most of them at scale---should default to
incomplete-block statistics such as Skillings--Mack rather than silently
discarding blocks.

Second, our LLM judge is substantially reliable for factual verification
(hallucination $\kappa=0.640$) and near-useless for architectural judgment
($\kappa=0.177$). Given the rapid adoption of model-based
evaluation~\cite{zheng2023judging,gu2025survey}, per-criterion calibration
against human raters should be reported rather than assumed. The near-chance
inter-human agreement on security-test relevance ($\kappa=0.059$) further
suggests that some evaluation criteria may not be reliably measurable by humans
either, and that disagreement should be reported rather than adjudicated away.

\section{Threats to Validity}
\label{sec:threats}

\textbf{Construct.} The human corpus is \emph{not} a matched control: those
templates were not written against our scenarios, and many are curated examples
rather than production infrastructure. Example templates may be deliberately
minimal, which could bias the baseline in either direction; we do not claim to
know which. Vendor reasoning mechanisms also differ (token budget versus effort
level) and are not pooled.

A second construct limitation concerns the prompts themselves, and we state it
plainly because an earlier version of this work got it wrong. A majority of
scenarios prescribe a security state rather than specifying function alone
(Section~\ref{sec:promptclass}). Findings on those scenarios partly measure
instruction-following rather than unprompted default posture. This was
identified by an independent audit of our released
artifacts~\cite{iacguardv}, reproduced exactly against our own data, and has been
corrected throughout; the first preprint of this paper described the prompts as
functional-only, which they are not. Prompt class is additionally confounded
with provider, format, and complexity, and the human corpus is not
prompt-matched, so the class-stratified figures in
Table~\ref{tab:promptclass} are descriptive rather than corrected causal
estimates. We report them, and the disagreement between two independent
classifications, instead of selecting whichever scheme is most favourable.

\textbf{Internal.} Scanner findings indicate policy deviations, not proven
exploitability; ``zero findings'' means only that three rulesets flagged nothing.
Checkov does not emit severity tiers without a commercial subscription, so
severity analysis reflects Trivy and KICS only. Scenarios were authored by a
model that also appears under test, an unquantified residual risk mitigated by
requirements-only phrasing and independent human and judge review.

\textbf{Measurement infrastructure.} During this study we identified several
defects that produced plausible-looking but incorrect results with no error
raised: a platform encoding mismatch that aborted scanning for an entire
configuration; a filename-handling rule that silently discarded one engine's
reports for configurations whose names contain a period; a schema mismatch that
caused one engine's findings to parse as zero for ten of twelve configurations;
a resource-count source that degraded to a constant denominator, inflating an
apparent effect to $+2528\%$ where the corrected value is $-14\%$; and
ungenerated scenarios being scored as zero-vulnerability. Each would have
survived review because the resulting numbers looked reasonable. We report this
explicitly: in multi-tool benchmarks, derived counts must be reconciled against
source artifacts rather than trusted, and we release coverage manifests and
regeneration scripts so that any published figure can be recomputed.

\textbf{External.} Hosted models change without notice; results describe the
configurations as accessed. Terraform dominates the corpus, and engine rule
coverage differs by format. Statistical power for the paired reasoning contrasts
is limited ($n\leq60$), and only two model families expose reasoning arms.

\section{Conclusion and Future Work}
\label{sec:conclusion}

Measured against human-authored infrastructure through an identical toolchain
and matched on artifact size, current language models produce IaC with
approximately $3.2\times$--$3.9\times$ the vulnerability density of human
engineers, with remarkable consistency across vendors and weight availability.
The gap is widest on the simplest tasks. Vendor reasoning modes deliver a real
but modest improvement and meaningfully outperform prompted chain-of-thought,
which alone provides no measurable security benefit---though the mechanism is
barely engaged on this task class, consuming under $1\%$ of output tokens.

Future work should pursue two controls in particular. The first is a
purpose-built functional-only prompt set: the present corpus cannot cleanly
separate unprompted default posture from instruction-following, because its
functional-only stratum is small and classification-sensitive
(Section~
ef{sec:promptclass}). The second is a matched human control authored against the same
scenarios; larger samples and additional model families for the reasoning
contrasts; a sensitivity sweep over reasoning-budget settings; exploitability
triage to connect policy findings to realizable attack paths; deployment-time
validation beyond static analysis; and improved rubric anchoring, given that
expert agreement on security-test relevance was near chance.

\section*{Acknowledgements}
We thank the three practicing cloud and security engineers who independently
reviewed the stratified scenario sample that underpins
Section~\ref{sec:human-eval}. They scored every sampled scenario without seeing
one another's ratings, and their disagreement---not merely their
agreement---shaped the reliability analysis and the limits we place on the LLM
judge. All three consented to be named:

\begin{itemize}\raggedright
  \item \textbf{Abhishek Pandey} ---
    \href{mailto:Abhishekp15b@iimk.edu.in}{Abhishekp15b@iimk.edu.in} ---
    \href{https://www.linkedin.com/in/abhishek-pandey-52926419/}{linkedin.com/in/abhishek-pandey-52926419}
  \item \textbf{Manickam Venkatachalam} ---
    \href{mailto:dev.venkey2000@gmail.com}{dev.venkey2000@gmail.com} ---
    \href{https://www.linkedin.com/in/manickam-venkatachalam/}{linkedin.com/in/manickam-venkatachalam}
  \item \textbf{Prajjuwal Varshney} ---
    \href{mailto:prajjuwalvarshney@gmail.com}{prajjuwalvarshney@gmail.com} ---
    \href{https://in.linkedin.com/in/prajjuwal-varshney-42861a107}{linkedin.com/in/prajjuwal-varshney-42861a107}
\end{itemize}

\noindent Their raw scores are released in anonymized form (\texttt{R1}--\texttt{R3});
the mapping between identifiers and identities is retained only by the author and
is not distributed.

We further thank \textbf{Lokesh Chauhan} (author of
IaC-Guard-V~\cite{iacguardv}, QRS 2026), who independently reproduced this
benchmark's released artifacts and identified that a majority of the scenario
prompts prescribe a security state, contrary to how the first preprint described
them. The finding reproduced exactly against our own data and materially improved
the paper's interpretation; Section~\ref{sec:promptclass} and the
threats-to-validity discussion exist because of it. Raising it with us directly,
rather than through review, was a generous act of scholarship.

\section*{Data Availability}
The pipeline, statistical analysis code, figure generation, coverage manifests,
and threats-to-validity documentation are released. Every figure and table is
regenerated from the released data by a single command. The generated corpus,
raw scanner output, human reference corpus, and all derived result tables are
archived at
\href{https://huggingface.co/datasets/AnimeshShaw/GenIaC-SecBench}{huggingface.co/datasets/AnimeshShaw/GenIaC-SecBench};
the code is at
\href{https://github.com/AnimeshShaw/GenIaC-SecBench}{github.com/AnimeshShaw/GenIaC-SecBench}.

\bibliographystyle{IEEEtran}
\bibliography{references}

\begin{thebibliography}{10}
\providecommand{\url}[1]{#1}
\csname url@samestyle\endcsname
\providecommand{\newblock}{\relax}
\providecommand{\bibinfo}[2]{#2}
\providecommand{\BIBentrySTDinterwordspacing}{\spaceskip=0pt\relax}
\providecommand{\BIBentryALTinterwordstretchfactor}{4}
\providecommand{\BIBentryALTinterwordspacing}{\spaceskip=\fontdimen2\font plus
\BIBentryALTinterwordstretchfactor\fontdimen3\font minus
  \fontdimen4\font\relax}
\providecommand{\BIBforeignlanguage}[2]{{%
\expandafter\ifx\csname l@#1\endcsname\relax
\typeout{** WARNING: IEEEtran.bst: No hyphenation pattern has been}%
\typeout{** loaded for the language `#1'. Using the pattern for}%
\typeout{** the default language instead.}%
\else
\language=\csname l@#1\endcsname
\fi
#2}}
\providecommand{\BIBdecl}{\relax}
\BIBdecl

\bibitem{rahman2019seven}
A.~Rahman, C.~Parnin, and L.~Williams, ``The seven sins: Security smells in
  infrastructure as code scripts,'' in \emph{IEEE/ACM International Conference
  on Software Engineering (ICSE)}, 2019, pp. 164--175.

\bibitem{verdet2023exploring}
A.~Verdet, M.~Hamdaqa, L.~Da~Silva, and F.~Khomh, ``Exploring security
  practices in infrastructure as code: An empirical study,'' \emph{Empirical
  Software Engineering}, 2023.

\bibitem{pearce2022asleep}
H.~Pearce, B.~Ahmad, B.~Tan, B.~Dolan-Gavitt, and R.~Karri, ``Asleep at the
  keyboard? assessing the security of {GitHub} {Copilot}'s code
  contributions,'' in \emph{IEEE Symposium on Security and Privacy (S\&P)},
  2022, pp. 754--768.

\bibitem{sandoval2023lost}
G.~Sandoval, H.~Pearce, T.~Nys, R.~Karri, S.~Garg, and B.~Dolan-Gavitt, ``Lost
  at {C}: A user study on the security implications of large language model
  code assistants,'' in \emph{USENIX Security Symposium}, 2023, pp. 2205--2222.

\bibitem{tihanyi2025secure}
N.~Tihanyi, T.~Bisztray, M.~A. Ferrag, R.~Jain, and L.~C. Cordeiro, ``How
  secure is {AI}-generated code: A large-scale comparison of large language
  models,'' \emph{Empirical Software Engineering}, 2025.

\bibitem{perry2023users}
N.~Perry, M.~Srivastava, D.~Kumar, and D.~Boneh, ``Do users write more insecure
  code with {AI} assistants?'' in \emph{ACM SIGSAC Conference on Computer and
  Communications Security (CCS)}, 2023, pp. 2785--2799.

\bibitem{vargas2026securityfirst}
G.~Vargas, R.~B. Mansilha, and D.~Kreutz, ``Security-first evaluation of
  text-to-{Terraform} generation,'' \emph{arXiv preprint arXiv:2608.02672},
  2026, accepted at SBSeg 2026.

\bibitem{skillings1981distribution}
J.~H. Skillings and G.~A. Mack, ``On the use of a {Friedman}-type statistic in
  balanced and unbalanced block designs,'' \emph{Technometrics}, vol.~23,
  no.~2, pp. 171--177, 1981.

\bibitem{siddiq2022securityeval}
M.~L. Siddiq and J.~C.~S. Santos, ``{SecurityEval} dataset: Mining
  vulnerability examples to evaluate machine learning-based code generation
  techniques,'' in \emph{International Workshop on Mining Software Repositories
  Applications for Privacy and Security (MSR4P\&S)}, 2022, pp. 29--33.

\bibitem{liu2023your}
J.~Liu, C.~S. Xia, Y.~Wang, and L.~Zhang, ``Is your code generated by {ChatGPT}
  really correct? rigorous evaluation of large language models for code
  generation,'' \emph{Advances in Neural Information Processing Systems
  (NeurIPS)}, 2023.

\bibitem{saavedra2023glitch}
N.~Saavedra and J.~F. Ferreira, ``{GLITCH}: Automated polyglot security smell
  detection in infrastructure as code,'' in \emph{IEEE/ACM International
  Conference on Automated Software Engineering (ASE)}, 2023.

\bibitem{rahman2019systematic}
A.~Rahman, R.~Mahdavi-Hezaveh, and L.~Williams, ``A systematic mapping study of
  infrastructure as code research,'' \emph{Information and Software
  Technology}, vol. 108, pp. 65--77, 2019.

\bibitem{checkov}
{Prisma Cloud / Bridgecrew}, ``Checkov: Policy-as-code for infrastructure as
  code,'' \url{https://github.com/bridgecrewio/checkov}, 2026.

\bibitem{trivy}
{Aqua Security}, ``Trivy: Comprehensive security scanner,''
  \url{https://github.com/aquasecurity/trivy}, 2026.

\bibitem{kics}
{Checkmarx}, ``{KICS}: Keeping infrastructure as code secure,''
  \url{https://github.com/Checkmarx/kics}, 2026.

\bibitem{cis_benchmarks}
{Center for Internet Security}, ``{CIS} benchmarks,''
  \url{https://www.cisecurity.org/cis-benchmarks}, 2026.

\bibitem{wei2022chain}
J.~Wei, X.~Wang, D.~Schuurmans, M.~Bosma, B.~Ichter, F.~Xia, E.~Chi, Q.~Le, and
  D.~Zhou, ``Chain-of-thought prompting elicits reasoning in large language
  models,'' \emph{Advances in Neural Information Processing Systems (NeurIPS)},
  2022.

\bibitem{kojima2022large}
T.~Kojima, S.~S. Gu, M.~Reid, Y.~Matsuo, and Y.~Iwasawa, ``Large language
  models are zero-shot reasoners,'' \emph{Advances in Neural Information
  Processing Systems (NeurIPS)}, 2022.

\bibitem{sprague2025cot}
Z.~Sprague, F.~Yin, J.~D. Rodriguez \emph{et~al.}, ``To {CoT} or not to {CoT}?
  chain-of-thought helps mainly on math and symbolic reasoning,''
  \emph{International Conference on Learning Representations (ICLR)}, 2025.

\bibitem{openai2024o1}
{OpenAI}, ``Learning to reason with {LLMs},'' \emph{OpenAI Technical Report},
  2024.

\bibitem{deepseek2025r1}
{DeepSeek-AI}, ``{DeepSeek-R1}: Incentivizing reasoning capability in {LLMs}
  via reinforcement learning,'' \emph{arXiv preprint arXiv:2501.12948}, 2025.

\bibitem{anthropic2025thinking}
{Anthropic}, ``Extended thinking,''
  \url{https://docs.anthropic.com/en/docs/build-with-claude/extended-thinking},
  2026.

\bibitem{zheng2023judging}
L.~Zheng, W.-L. Chiang, Y.~Sheng \emph{et~al.}, ``Judging {LLM}-as-a-judge with
  {MT-Bench} and {Chatbot Arena},'' \emph{Advances in Neural Information
  Processing Systems (NeurIPS)}, 2023.

\bibitem{gu2025survey}
J.~Gu, X.~Jiang, Z.~Shi \emph{et~al.}, ``A survey on {LLM}-as-a-judge,''
  \emph{arXiv preprint arXiv:2411.15594}, 2025.

\bibitem{panickssery2024llm}
A.~Panickssery, S.~R. Bowman, and S.~Feng, ``{LLM} evaluators recognize and
  favor their own generations,'' \emph{Advances in Neural Information
  Processing Systems (NeurIPS)}, 2024.

\bibitem{ralph2021empirical}
P.~Ralph \emph{et~al.}, ``Empirical standards for software engineering
  research,'' \emph{arXiv preprint arXiv:2010.03525}, 2021.

\bibitem{friedman1937use}
M.~Friedman, ``The use of ranks to avoid the assumption of normality implicit
  in the analysis of variance,'' \emph{Journal of the American Statistical
  Association}, vol.~32, no. 200, pp. 675--701, 1937.

\bibitem{wilcoxon1945individual}
F.~Wilcoxon, ``Individual comparisons by ranking methods,'' \emph{Biometrics
  Bulletin}, vol.~1, no.~6, pp. 80--83, 1945.

\bibitem{holm1979simple}
S.~Holm, ``A simple sequentially rejective multiple test procedure,''
  \emph{Scandinavian Journal of Statistics}, vol.~6, no.~2, pp. 65--70, 1979.

\bibitem{demsar2006statistical}
J.~Dem{\v{s}}ar, ``Statistical comparisons of classifiers over multiple data
  sets,'' \emph{Journal of Machine Learning Research}, vol.~7, pp. 1--30, 2006.

\bibitem{liang1986longitudinal}
K.-Y. Liang and S.~L. Zeger, ``Longitudinal data analysis using generalized
  linear models,'' \emph{Biometrika}, vol.~73, no.~1, pp. 13--22, 1986.

\bibitem{hilbe2011negative}
J.~M. Hilbe, \emph{Negative Binomial Regression}, 2nd~ed.\hskip 1em plus 0.5em
  minus 0.4em\relax Cambridge University Press, 2011.

\bibitem{mann1947test}
H.~B. Mann and D.~R. Whitney, ``On a test of whether one of two random
  variables is stochastically larger than the other,'' \emph{Annals of
  Mathematical Statistics}, vol.~18, no.~1, pp. 50--60, 1947.

\bibitem{massey1951kolmogorov}
F.~J. Massey~Jr., ``The {Kolmogorov-Smirnov} test for goodness of fit,''
  \emph{Journal of the American Statistical Association}, vol.~46, no. 253, pp.
  68--78, 1951.

\bibitem{iacguardv}
\BIBentryALTinterwordspacing
L.~Chauhan, ``{IaC-Guard-V}: A fail-closed verification framework for
  {LLM}-generated infrastructure-as-code repairs,'' 2026, independent audit of
  the GenIaC-SecBench released artifacts. Accompanying paper at QRS 2026,
  Springer LNCS. [Online]. Available:
  \url{https://github.com/lokesh0186/iac-guard-v}
\BIBentrySTDinterwordspacing

\bibitem{fleiss1971measuring}
J.~L. Fleiss, ``Measuring nominal scale agreement among many raters,''
  \emph{Psychological Bulletin}, vol.~76, no.~5, pp. 378--382, 1971.

\bibitem{landis1977measurement}
J.~R. Landis and G.~G. Koch, ``The measurement of observer agreement for
  categorical data,'' \emph{Biometrics}, vol.~33, no.~1, pp. 159--174, 1977.

\bibitem{cohen1960coefficient}
J.~Cohen, ``A coefficient of agreement for nominal scales,'' \emph{Educational
  and Psychological Measurement}, vol.~20, no.~1, pp. 37--46, 1960.

\end{thebibliography}

\end{document}